\documentclass[useAMS,usenatbib]{mn2e}
\usepackage{graphics,latexsym,epsfig,graphicx,amssymb,lscape,aas_macros}     

\renewcommand{\d}[0]{{\rm d}}
\newcommand{\ave}[1]{\langle #1 \rangle}
\newcommand{\Ave}[1]{\Big\langle #1 \Big\rangle}
\newcommand{\Ref}[1]{\ref{#1}}

\renewcommand{\vec}[1]{{\bmath{#1}}}
\newcommand{\mat}[1]{\mathbfss{#1}}

\newcommand{\msol}[0]{{\rm M}_\odot}
\newcommand{\high}[1]{#1}
\newcommand{\changed}[1]{#1}
\title[3-D mapping with 3-D weak lensing]
  {Unfolding the matter distribution using 3-D weak gravitational lensing}
\author[Simon, Taylor~\&~Hartlap]
  {P.~Simon$^1$, A.N. Taylor$^1$ and J. Hartlap$^2$\\
  $^1$The Scottish Universities Physics Alliance (SUPA), Institute for
  Astronomy, School of Physics, \\University of Edinburgh, Royal
  Observatory, Blackford Hill, Edinburgh EH9 3HJ, UK\\
  $^2$Argelander-Institut f\"ur Astronomie, Universit\"at Bonn, Auf
  dem H\"ugel 71, D-53121 Bonn, Germany}
  
\date{Released 2007 Xxxxx XX}

\pagerange{\pageref{firstpage}--\pageref{lastpage}} \pubyear{2008}

\def\LaTeX{L\kern-.36em\raise.3ex\hbox{a}\kern-.15em
    T\kern-.1667em\lower.7ex\hbox{E}\kern-.125emX}

\begin{document}

\maketitle

\begin{abstract}
  
  Combining redshift and galaxy shape information offers new exciting
  ways of exploiting the gravitational lensing effect for studying the
  large scales of the cosmos. One application is the three-dimensional
  reconstruction of the matter density distribution which is explored
  in this paper. We give a generalisation of an already known
  minimum-variance estimator of the 3-D matter density distribution
  that facilitates the combination of thin redshift slices of sources
  with samples of broad redshift distributions for an optimal
  reconstruction; sources can be given individual statistical
  weights. We show how, in principle, intrinsic alignments of source
  ellipticities or shear/intrinsic alignment correlations can be
  accommodated, albeit these effects are not the focus of this
  paper. We describe an efficient and fast way to implement the
  estimator on a contemporary desktop computer. Analytic estimates for
  the noise and biases in the reconstruction are given.  Some
  regularisation (Wiener filtering) of the estimator, adjustable by a
  tuning parameter, is necessary to increase the signal-to-noise to a
  sensible level and to suppress oscillations in radial
  direction. This, however, introduces as side effect a systematic
  shift and stretch of structures in radial direction. This bias can
  be expressed in terms of a radial PSF, comprising the limitations of
  the reconstruction due to source shot-noise and the unavoidably
  broad lensing kernel. \changed{We conclude that a 3-D mass-density
    reconstruction on galaxy cluster scales ($\sim1\,\rm Mpc$) is
    feasible but, for foreseeable surveys, a map with a ${\rm
      S/N}\gtrsim3$ threshold is limited to structures with
    $M_{200}\gtrsim1\times10^{14}\msol h^{-1}$, or
    $7\times10^{14}\,\msol h^{-1}$, at low to moderate redshifts
    ($z=0.1$ or $0.6$). However, we find that a heavily smoothed
    full-sky map of the very large-scale density field may also be
    possible as the S/N of reconstructed modes increases towards
    larger scales.}  \changed{Future improvements of the method may be
    obtained by including higher-order lensing information (flexion)
    which could also be implemented into our algorithm.}
\end{abstract}


%-------------------------------------------------------------------
%\tableofcontents

\section{Introduction}
\label{firstpage}

Since its detection about a decade ago
\citep{2000MNRAS.318..625B,2000astro.ph..3338K,2000A&A...358...30V,2000Natur.405..143W}
weak gravitational lensing, or cosmic shear -- the small, coherent
distortion of distant galaxy images due to the intervening
large-scale matter distribution in the cosmos -- has become a
well-established tool for studying the dark Universe \citep[for
recent reviews
see:][]{2008arXiv0805.0139H,2006glsw.conf..269S,2006astro.ph.12667M,2003astro.ph..5089V}.
The phenomena of gravitationally scattering of light by massive
structures is well described by General Relativity. Hence lensing
provides us with a direct probe of the mass-distribution in the
Universe. The spectrum of applications for exploiting this pool of
information includes the interpretation of correlations of the
cosmic shear signal to constrain cosmological parameters, the
cross-correlation of cosmic shear with galaxy positions for
studying the typical matter environment of galaxies or galaxy
biasing, and the fascinating possibility of mapping the
large-scale mass distribution in the Universe.

Provided that General Relativity itself is accurate
\citep{2007PhRvD..75b3519H,2001PhRvD..64h3004U} the statistical
analysis of the weak shear signal for cosmological parameter
estimation provides us with ample indirect evidence that most of the
matter of the Universe is not made up of familiar baryonic matter, but
of so-called Dark Matter. Moreover the largest piece in the energy
budget in the present Universe is occupied by the so-called Dark
Energy which seems to be accelerating the expansion of the Universe
\citep[see e.g.][ for a recent
review]{2006ewg3.rept.....P,2006astro.ph..9591A}. As the area, $A$, of
weak lensing surveys grows, the statistical uncertainty on the
parameters of this cosmological model will decrease as $1/\sqrt{A}$,
allowing even General Relativity itself to be tested. The main issue
for lensing surveys for extracting cosmological parameters will be the
control of systematics at the 1\%-level \citep{2006MNRAS.366..101H}.

As well as statistical analysis, weak lensing can also be used to
directly visualise the matter distribution in the
Universe. \citet{1993ApJ...404..441K} first showed how to map the
projected 2-D mass distribution from weak
lensing. \citet{2001astro.ph.11605T}, \citet{2003MNRAS.344.1307B} and
\citet{2004MNRAS.353.1176T} showed how to extend this to 3-D mapping
by combining shear data with redshift distances (3-D weak
lensing). Visualisation of the Dark Matter distribution has the clear
advantage that one can directly compare the matter distribution with
that of galaxies and gas \citep[e.g.][]{2006ApJ...648L.109C}.

For a given resolution scale, mass mapping does not benefit from a
larger area. The noise of a mass map scales as
$\sigma_\epsilon/\sqrt{N}$ \citep{1993ApJ...404..441K,
2002PhRvD..66f3506H} where $N=n \theta_s$ is the surface number of
lensing relevant galaxies (sources) within a resolved area,
$\theta_s$, the mean surface density of galaxies is $n$, and
$\sigma_\epsilon$ is the intrinsic random ellipticity distribution of
sources. It is unlikely that $n$ and $\sigma_\epsilon$ will change
compared to current surveys as dramatically as the survey area, while
the uncertainty in redshifts must also be added for 3-D mass mapping.

Ambitious survey campaigns are either under way (e.g. \mbox{CFHTLS},
\mbox{RCS2}, \mbox{COSMOS}), about to commence (e.g.
\mbox{Pan-STARRS}, \mbox{VST-KIDS}, \mbox{DES}) to cover larger areas
of sky, while surveys being planned for the near future
(e.g. \mbox{LSST}, \mbox{Euclid}) will map the full sky to great
depths and high resolution. This will provide the lensing community
with large galaxy catalogues with billions of galaxies including
multi-colour information for photometric redshifts and accurate galaxy
image measurements. These surveys mark the beginning of a new era for
lensing as the availability of a large set of sources with redshifts
and multi-colour data will allow us to accurately probe the evolution
of mass-clustering as a function of cosmic time.

This paper addresses in detail the question of how to optimally
reconstruct the \emph{3-D matter density} solely based on a lensing
catalogue with redshift information. \changed{Applying Wiener
filtering to cosmic shear follows in the tradition of other successful
cosmological applications, initially outlined for astrophysical
applications by \citet{1992ApJ...398..169R} and applied to galaxy
redshift surveys \citep{1994ApJ...423L..93L, 1995ApJ...449..446Z,
1997MNRAS.287..425W}, CMB maps \citep{1994ApJ...432L..75B,
1999MNRAS.302..663B} and peculiar velocity fields
\citep{1999ApJ...520..413Z}.} Here we describe a linear
minimum-variance reconstruction operator which is a generalisation of
the Wiener filter method proposed by
\citet{2002PhRvD..66f3506H}. \citet{2003MNRAS.344.1307B} also proposed
a Wiener filtering method, which was first applied in
\citet{2004MNRAS.353.1176T} for a 3-D reconstruction of the COMBO-17
survey \citep{2004A&A...421..913W} and in \citet{2007Natur.445..286M}
to the COSMOS survey. However this was to reconstruct the
gravitational potential, rather than the mass-density field, which
improved the noise properties of the 3-D reconstruction on cluster
scales \citep[see][]{2003MNRAS.344.1307B}. \citet{2002PhRvD..66f3506H}
pointed out that on larger scales the signal-to-noise improves
allowing density-reconstruction. Both \citet{2002PhRvD..66f3506H} and
\citet{2003MNRAS.344.1307B} proposed reconstructing the 3-D
convergence field in a first step, and then applying the Wiener filter
along each individual line-of-sight, ignoring the correlation of
errors between different line-of-sights (radial mapmaking). In the
application to the COMBO-17 and COSMOS surveys, all of the redshift
bins where assumed to have the same noise level, which is sub-optimal.

In this paper, the transition from 3-D shear to lens-plane density
contrast is done in one step, along all line-of-sights
simultaneously, including all correlations between different
lines-of-sight and lens planes. Another extension consists in
introducing an additional parameter that regulates the impact of
the Wiener regularisation. With the approach outlined here we show
how all available redshift information, i.e. redshift slices and
sub-samples with broad redshift distributions and individual
statistical weights of sources, can be combined in one framework.
The primary focus of this study will be the signal-to-noise and
biases that are introduced in the 3-D reconstruction if we
regularise the process with a prior. Regularisation will be
unavoidable due to the poor signal-to-noise of a 3-D density map
based on lensing alone. Besides the theoretical considerations, we
also describe how the reconstruction algorithm can be efficiently
implemented in practise.

This paper is laid out as follows. We start off in Sect.
\ref{sect:mlinv} with a very general discussion of inverse
estimators for linear problems. Sect. \ref{sect:method} continues
by specifying the problem of 3-D reconstructions in two steps: the
convergence tomography and the lens-plane matter density contrast
from the cosmic shear tomography. Both problems fall into the
category of linear inversion problems. Sect. \ref{sect:fourier}
discusses the signal-to-noise of mass density reconstructions, the
main focus of this paper, and biases therein that are introduced
by a prior.  As prior we consider expected density correlations in
radial or transverse direction. The practical implementation of
the estimators, Appendix \ref{sect:implementation}, requires some
optimisation, because a naive implementation would force us to
handle extremely large matrices even on relatively small grids.
Finally in Sect. \ref{sect:mockdata}, we apply the algorithm to a
fictive, simulated survey with specification we should expect for
a weak lensing surveys within the next twenty years.

We use as fiducial cosmology a $\Lambda$CDM model (adiabatic
fluctuations) with \mbox{$\Omega_{\rm m}=0.24$},
\mbox{$\Omega_{\rm b}=0.0416$},
\mbox{$\Omega_{\Lambda}=1-\Omega_{\rm m}$} and
\mbox{$H_{0}=h\,100\,{\rm km}\,{\rm s}^{-1}\,{\rm Mpc}^{-1}$} with
\mbox{$h=0.732$}. The normalisation of the matter fluctuations
within a sphere of radius $8\,h^{-1}\rm Mpc$ at redshift zero is
assumed to be $\sigma_8=0.76$. For the spectral index of the
primordial matter power spectrum we use $n_{\rm s}=0.96$. These
values are consistent with the third-year WMAP results
\citep{2007ApJS..170..377S}.

\section{General linear inversions}
\label{sect:mlinv}

\subsection{Estimators for linear problems}
All problems discussed in this paper are linear inversion
problems. For such problems, a vector of observables, $\vec{d}$, is
known to be a linear transformation of a certain vector of underlying
parameters, $\vec{s}$:
\begin{equation}\label{invgeneral}
  \vec{d}=\mat{R}\vec{s}+\vec{n}\;,
\end{equation}
where $\mat{R}$ is a linear projection from $\vec{s}$-space to
$\vec{d}$-space. The vector $\vec{n}$ denotes a set of random
variables that introduces noise to the observables.  The statistical
uncertainties of the observables are thought to be vanishing on
average, $\ave{\vec{n}}=\vec{0}$, and may be a function of
$\vec{s}$. All involved matrices and vectors may be complex numbers,
the dagger$^\dagger$ denotes the complex conjugate transpose.

The inversion of Eq. \Ref{invgeneral} with respect to $\vec{s}$ for a
given $\vec{d}$ can, in general, only be done in a probabilistic way
because of a) the noise in the observation and b) because $\mat{R}$ is
usually not regular or not even a square matrix. One approach, the
maximum-likelihood (ML) approach, uses the posterior likelihood
function for $\vec{s}$ given the observation $\vec{d}$ and known
probability distributions of the involved random variables to find the
most likely solution for $\vec{s}$
\citep[e.g.][]{2008MNRAS.tmp..895K}:
\begin{equation}
  P(\vec{s}|\vec{d})=
  \frac{P(\vec{d}|\vec{s})P(\vec{s})}{P(\vec{d})}\propto
  P(\vec{d}|\vec{s})P(\vec{s})\;.
\end{equation}
The Bayesian evidence $P(\vec{d})$ is a constant with no interest
for the scope of this paper, $P(\vec{s})$ is a prior on $\vec{s}$
and $P(\vec{d}|\vec{s})$ the likelihood function of $\vec{d}$ given
some parameters $\vec{s}$. 

In a weaker approach to the inversion problem, one can demand to find
a linear operator, $\mat{H}\vec{d}$ say, acting on the observation
$\vec{d}$ that minimises the variance of residuals between the guessed
signal and true signal,
\mbox{$\ave{(\vec{s}-\mat{H}\mat{d})(\vec{s}-\mat{H}\mat{d})^\dagger}$},
without the strong assumption of $P(\vec{s})$ and $P(\vec{s}|\vec{d})$
\citep{1998ApJ...506...64S}. All that is now required, are the
statistical means of $\vec{s}$ and $\vec{n}$, both assumed to be zero
in our case (or can be made zero by definition), and their covariances
\mat{$\mat{N}\equiv\ave{\vec{n}\vec{n}^\dagger}$} and
\mbox{$\mat{S}\equiv\ave{\vec{s}\vec{s}^\dagger}$}. Minimising the
covariance of the residuals for all conceivable noise and signal
configurations results in a minimum variance (MV) estimator:
\begin{eqnarray}
  \vec{s}_{\rm MV}=
  \mat{H}\vec{d}&=&\left[
    \mat{S}^{-1}+\mat{R}^\dagger\mat{N}^{-1}\mat{R}
    \right]^{-1}\,\mat{R}^\dagger\mat{N}^{-1}\vec{d}\\
  \label{mlinv}
  &=&\left[
    \mat{1}+\mat{S}\mat{R}^\dagger\mat{N}^{-1}\mat{R}
    \right]^{-1}\,\mat{S}\mat{R}^\dagger\mat{N}^{-1}\vec{d}\;,
\end{eqnarray}
where $\mat{1}$ is a unity matrix of the dimensions of $\mat{S}$. We
call this filter $\mat{H}$ a Wiener filter. Here, we have explicitly
assumed no correlation between signal and noise,
i.e. $\ave{\vec{s}\vec{n}^\dagger}=\mat{0}$ (see Appendix
\ref{sect:shearintr}).

If we would not like or cannot make strong assumptions on the signal
covariance, we can assume it to have infinite variance,
\mat{$\mat{S}^{-1}=\mat{0}$}, avoiding any preference in parameter
space and only using the noise covariance as weight (inverse variance filter):
\begin{equation}\label{mlinvnoprior}
  \vec{s}_{\rm IV}= \left[\mat{R}^\dagger\mat{N}^{-1}\mat{R}
    \right]^{-1}\,\mat{R}^\dagger\mat{N}^{-1}\vec{d}\;.
\end{equation}
In the Bayesian interpretation of the MV-estimator we
would not have any prior on $\vec{s}$.

For multivariate Gaussian distributions $P(\vec{s})$ and
$P(\vec{d}|\vec{s})$ the maximum-likelihood and minimum-variance
solution are identical, so that in the Gaussian case the most likely
solution is the minimum-variance solution. For this paper, the
reconstructed fields will be cosmological density fields smoothed to
large scales which makes them roughly Gaussian fields.

Commonly, the sizes of the vectors and matrices can become large so
that multiplications between vectors and matrices, a ${\cal O}(N^2)$
process, or matrix inversion, a ${\cal O}(N^3)$ process
\citep{1992nrca.book.....P}, can take a very long time or require
much computer memory if done without optimisations. The last step in
Eq. \Ref{mlinv} ensures that we have one matrix inversion less.

It may be that in the estimators Eq. \Ref{mlinv} or, in particular,
Eq. \Ref{mlinvnoprior} the matrices in the square brackets are not regular
and hence cannot be inverted. This happens in reconstructions of the
lensing convergence due to the mass-sheet degeneracy. This problem can
be solved by enforcing another regularisation that asserts the mean
over all elements of $\vec{s}_{\rm MV}$ to be \emph{exactly} a
particular constant $s_0$. This makes the minimisation problem of
finding $\mat{H}$ a minimisation with additional constraint
$\vec{s}^\dagger\vec{s}=s_0$ breaking the degeneracy
\citep{2002PhRvD..66f3506H}.

Alternatively, we can assert $\vec{s}^\dagger\vec{s}$ to be
\emph{close to} zero (or another constant value) confined by a
Gaussian probability distribution as prior $\propto{\rm
e}^{-\frac{1}{2}\lambda\vec{s}^\dagger\vec{s}}$; $\lambda$ is a
constant constraining the allowed regime of
$\vec{s}^\dagger\vec{s}$. This is achieved by adding another equation
to the minimisation of the aforementioned residual matrix for
$\mat{H}$ and results in the so-called Tikhonov regularisation
\citep{1996astro.ph.12006W} also breaking the degeneracy:
\begin{equation}\label{mlinvnoprior2}
  \vec{s}_{\rm MV}=
  \left[\lambda\mat{1}+\mat{R}^\dagger\mat{N}^{-1}\mat{R}
    \right]^{-1}\,\mat{R}^\dagger\mat{N}^{-1}\vec{d}\;.
\end{equation}
Note that using the Wiener filter, and therefore a constraint on the
allowed variance of $\vec{s}^\dagger\vec{s}$ via $\mat{S}$, usually
already does the job and breaks, in particular, the mass-sheet
degeneracy.

A fusion of a Wiener filter and a filter without prior, i.e.
$\mat{S}^{-1}=\mat{0}$, is introduced by a tuning parameter
(``$\alpha$-tuning''), $\alpha\in[0,\infty[$, for the Wiener filter
that changes the impact of the Wiener filter by scaling up the
expected signal-to-noise in the data by
$\mat{S}^{-1}\mapsto\alpha\mat{S}^{-1}$ \citep[Saskatoon
filter:][]{1997ApJ...480L..87T,1997ApJ...474L..77T}:
\begin{equation}\label{eq:shakatoon}
  \vec{s}_{\rm MV}=
  \left[\alpha\mat{1}+\mat{S}\mat{R}^\dagger\mat{N}^{-1}\mat{R}
    \right]^{-1}\,\mat{S}\mat{R}^\dagger\mat{N}^{-1}\vec{d}\;.
\end{equation}
For $\alpha=1$, the Saskatoon filter is identical to the normal Wiener
filter, and for $\alpha=0$ we would apply the filter without any
prior.  The Saskatoon filter will be of central importance for the
reconstruction technique outlined in this paper.

\subsection{Covariance}
The (noise) covariance of the estimator Eq. \Ref{eq:shakatoon} is found by
working out
\begin{equation}
  {\rm Cov}{(\vec{s}_{\rm MV})}=
  \ave{\vec{s}_{\rm MV}^{}\vec{s}_{\rm MV}^\dagger}-
  \ave{\vec{s}_{\rm MV}}\ave{\vec{s}_{\rm MV}^\dagger}\;,
\end{equation}
with $\ave{\ldots}$ being the ensemble average over all realisations
of the noise $\vec{n}$ but leaving the embedded signal unchanged. This
results in the expression
\begin{equation}\label{covariance}
  {\rm Cov}{(\vec{s}_{\rm MV})}=\mat{W}\mat{X}\mat{W}^\dagger\;.
\end{equation}
Here, \mbox{$\mat{X}\equiv
  \left[\mat{R}^\dagger\mat{N}^{-1}\mat{R}\right]^{-1}$} is the
  covariance of $\vec{s}_{\rm MV}$ for the no-prior case and
  \mbox{$\mat{W}\equiv\mat{S}[\mat{S}+\alpha\mat{X}]^{-1}$} the Wiener
  filter in $\mat{s}$-space. The covariance matrix has two limiting
  regimes. In the regime where the data is noise dominated,
  \mbox{$|\mat{S}\mat{X}^{-1}|\ll 1$}, it is approximately
  \mbox{$\mat{S}\mat{X}^{-1}\mat{S}$}. On the other hand, for data
  where the noise is small compared to the signal,
  \mbox{$|\mat{S}\mat{X}^{-1}|\gg1$}, one finds \mbox{$\mat{X}$},
  which is exactly the covariance of the MV-estimator without any
  Wiener regularisation.

The power of the signal that gets through the filter (on average),
\begin{equation}
  {\rm Pow}(\vec{s}_{\rm MV})=
  \ave{\vec{s}_{\rm MV}\vec{s}_{\rm MV}^\dagger}^\prime-
  \ave{\vec{s}_{\rm MV}}^\prime\ave{\vec{s}_{\rm MV}^\dagger}^\prime\;,
\end{equation}
obtained by taking the ensemble average $\ave{\ldots}^\prime$ over
all possible signal realisations while setting the noise
\mbox{$\vec{n}=0$}, is\footnote{It is assumed that the covariance of
  the true signal is indeed identical to the Wiener prior,
  i.e. \mbox{$\ave{\vec{s}\vec{s}^\dagger}^\prime=\mat{S}$}.}
\begin{equation}\label{scovariance}
  {\rm Pow}(\vec{s}_{\rm MV})=
  \mat{W}\mat{S}\mat{W}^\dagger\;.
\end{equation}
In the signal dominated regime, this is roughly $\mat{S}$, but in
the noise dominated regime it is well described by
\mbox{$\mat{S}\mat{X}^{-1}\mat{S}\mat{X}^{-1}\mat{S}$}. This shows
that a Wiener filter suppresses the signal if the signal-to-noise
becomes low.

\subsection{Bias}
We briefly address here the question if the MV-estimators are unbiased
estimators, i.e. whether $\ave{\vec{s}_{\rm
MV}}^\prime=\vec{s}$. Using the above equations and
$\ave{\vec{d}}^\prime=\mat{R}\vec{s}$ we find
\begin{equation}
  \ave{\vec{s}_{\rm MV}}^\prime=\mat{W}\vec{s}
\end{equation}
for the Wiener filter. For sufficiently small noise,
\mbox{$\mat{N}^{-1}\!\!\rightarrow\infty$} the Wiener filer
asymptotically approaches \mbox{$\mat{X}\mat{X}^{-1}$}, which is the
identity matrix if
\mbox{$\mat{R}^\dagger\mat{N}^{-1}\mat{R}=\mat{X}^{-1}$} is invertible
and undefined otherwise. We may call problems in which $\mat{X}^{-1}$
is invertible well-posed. Note that in the forgoing discussion of the
asymptotic estimator covariances we assumed a well-posed problem.

Therefore, for well-posed problems the Wiener filter is asymptotically
unbiased for an increasingly higher signal-to-noise, and the
non-Wiener filter is always unbiased (no prior). As we will see later,
the Wiener filter applies a smoothing to the reconstruction by
rescaling of low S/N-modes to some degree which explains why it is
not an unbiased estimator in general.

Non-well posed problems loose information when applying $\mat{R}$ to
the original signal. Therefore, even with no noise one cannot recover
the exact signal. Those problems can be made treatable, by introducing
the Tikhonov regularisation, if we make
\mbox{$\mat{X}^{-1}+\lambda\mat{1}$} invertible. This, however, will
also to some extent bias the estimator. One finds
\mbox{$\mat{W}=(\mat{X}^{-1}+\lambda\mat{1})^{-1}\mat{X}^{-1}$} is the
bias matrix in this case.

For a Wiener filter an original signal in one of the components of
$\vec{s}$ is, on average, in the reconstruction rescaled and spread
over the other vector components $\vec{s}_{\rm MV}$.  The $i$th column
of the (bias) matrix $\mat{W}$ tells us how the original signal $s_i$
is spread over the other vector elements of $\vec{s}_{\rm MV}$ in the
reconstruction.

The apparently obvious thing to remove the bias in the estimate would
be to multiply the MV-estimator by the inverse, if it exists, of the
bias matrix:
\begin{equation}
  \mat{W}^{-1}\vec{s}_{\rm MV}=
  \mat{X}\mat{R}^\dagger\mat{N}^{-1}(\mat{R}\vec{s}+\vec{n})=
  \vec{s}+\mat{X}\mat{R}^\dagger\mat{N}^{-1}\vec{n}\;.
\end{equation}
This, however, yields exactly the estimator Eq. \Ref{mlinvnoprior},
removing the benefits of the Wiener filter. The noise covariance is
now $\mat{X}$ again as can be seen from the second term in the
r.h.s. of the last equation.

The Saskatoon filter, Eq. \Ref{eq:shakatoon}, is a compromise between
the advantages of the Wiener filter, Eq. \Ref{mlinv} (improved
signal-to-noise but biased), and the no-prior filter,
Eq. \Ref{mlinvnoprior} (noisy but unbiased). However, less Wiener
filtering means less smoothing in the, then noisier, reconstruction
which here will require an additional (non-optimal) smoothing after
filtering.

\citet{2002MNRAS.331..901Z} derive another linear filter, a specific
quadratic estimator, for unbiased reconstruction of the large-scale
matter distribution from galaxy redshift surveys. Principally, this
filter cold be employed in the context of this paper as well. However,
as it turns out, the UMV-estimator of \citet{2002MNRAS.331..901Z} is
identical to Eq. \Ref{eq:shakatoon} with setting $\mat{N}=\mat{1}$
(equal weighing of all data points) and $\alpha=0$, which is the
no-prior filter, Eq. \Ref{mlinvnoprior} with $\mat{N}=\mat{1}$ and is
therefore automatically included in the following
discussion. Moreover, as already pointed out, the no-prior filter is
singular for lensing applications so that some regularisation, such as
$\alpha>0$ or $\lambda>0$, would be required for the UMV-filter, then
being a biased as well.

\section{Reconstruction method}
\label{sect:method}

\subsection{Lensing convergence tomography} 

Let us see how this fits into the context of three-dimensional mass
reconstructions based on cosmic shear tomography.  In weak
gravitational lensing \citep[e.g.][]{2001PhR...340..291B}, the
observables are complex ellipticities, $E^{(i)}_k$ with $k$ being a
galaxy index, of galaxy sources that, for our purposes, belong to one
out of $i=1\ldots N_z$ different redshift bins. The source ellipticity
is, in the weak lensing regime, an unbiased estimator of the cosmic
shear, $\gamma$, at the position, $\vec{\theta}_k$, of the source. An
angular position on the sky is denoted by $\vec{\theta}$ which, for
convenience, is a two-dimensional Cartesian vector expressed as
complex number.

For the following, we assume that the vector
\begin{equation}
  \vec{\gamma}^{(i)}=
  \left(\epsilon^{(i)}_1\,\,\epsilon^{(i)}_2\,\,\ldots\,\,\epsilon^{(i)}_{N_{\rm
  g}}\right)^{\rm t}
\end{equation}
contains the complex ellipticities binned on a grid, $N_{\rm g}$ is
the number of grid points. Only sources belonging to the same redshift
bin are binned together onto an angular grid.  All source
ellipticities, $E^{(i)}_j$, contained inside the $k$th grid pixel and
the same redshift bin, $i$, are combined to one average ellipticity
\begin{equation}
  \epsilon^{(i)}_k\equiv
  \frac{\sum_jw^{(i)}_jE^{(i)}_j}{\sum_jw^{(i)}_j}\;,
\end{equation}
where $E^{(i)}_j$ and $w^{(i)}_j$ are the ellipticity and statistical
weight of the sources associated with the $k$th grid pixel,
respectively. If there are no sources inside the pixel or the sum of
all $w^{(i)}_j$ is vanishing, set $\epsilon^{(i)}_k=0$.  We will
further assume that sources of all redshift bins are binned onto the
same angular grid so that same elements $\epsilon^{(i)}_k$ and
$\epsilon^{(j)}_k$ of different bins $i$ and $j$ lie along the same
line-of-sight, $\vec{\theta}_k$.  By $A$ we denote the solid angle of
the grid pixels. Naturally, the number of grid points is identical for
all redshift bins.

The affiliation of a source to the $i$th redshift bin means that we
have prior knowledge on its redshift. We denote this prior
probability on the redshift, $z$, by $p_z^{(i)}(z)$. Although we are
using the term ``redshift bin'' for $p_z^{(i)}(z)$ in this context,
we would like to point out that the ``bins'' may be arbitrary, including
overlapping, distributions of sources in redshift. It may be
conceivable, for example, that only a small fraction of sources have
redshift information and can thereby be ordered into non-overlapping
z-bins. The cosmic shear information of all the other sources could
then be added as well by putting them into one ``bin'' with a wide
redshift distribution, which is overlapping with all the other
bins. This way, we can combine all available shear catalogues to
acquire the best possible reconstruction.\footnote{Note that the
  following assumes that the noise covariance is diagonal. If we
  have the same galaxies belonging to different bins this is only a
  good approximation if their fraction is small. If not, we have to
  account for this in the noise covariance.}

In the weak lensing regime, the lensing convergence
$\vec{\kappa}^{(i)}$ -- also arranged as $N_{\rm g}$-dimensional
vector of convergences at position $\vec{\theta}_k$ analogues to
$\vec{\gamma}^{(i)}$ -- is related to the source ellipticities by a
linear transformation $\hat{\mat{P}}_{\gamma\kappa}$, a complex
\mbox{$N_{\rm g}\times N_{\rm g}$} matrix \citep{2002PhRvD..66f3506H}:
\begin{equation}\label{transform}
  \vec{\gamma}^{(i)}=
  \hat{\mat{P}}_{\gamma\kappa}\vec{\kappa}^{(i)}+\vec{n}^{(i)}_\gamma\;,
\end{equation}
where $\vec{n}^{(i)}_\gamma$ is a vector of random intrinsic source
ellipticities. The linear transformation
$\hat{\mat{P}}_{\gamma\kappa}$ is for \mbox{$k\ne l$}
\citep{1993ApJ...404..441K}
\begin{equation}\label{pdef}
  [\hat{\mat{P}}_{\gamma\kappa}]_{kl}=
  -\frac{A}{\pi}\frac{1}{[\theta_{kl}^\ast]^2}\;,
\end{equation}
where $\theta_{kl}^\ast$ is the complex conjugate of the separation
vector $\vec{\theta}_l-\vec{\theta}_k$. For \mbox{$k=l$}, we set
$[\hat{\mat{P}}_{\gamma\kappa}]_{kl}=0$.

Note that, so far, we have carefully separated the different
redshift bins from each other by using the superscript
``$^{(i)}$''. For each bin $i$ there is one Eq. \Ref{transform}. We
can combine everything to one compact system of equations,
\begin{equation}
  \vec{\gamma}= \mat{P}_{\gamma\kappa}\vec{\kappa}+\vec{n}_\gamma\;,
\end{equation}
by putting together all information into the vectors:
\begin{eqnarray}
  \vec{\gamma}&\equiv&
  \left(\vec{\gamma}^{(1)},\vec{\gamma}^{(2)},\ldots,\vec{\gamma}^{(N_z)}\right)\;,\\
  \vec{\kappa}&\equiv&
  \left(\vec{\kappa}^{(1)},\vec{\kappa}^{(2)},\ldots,\vec{\kappa}^{(N_z)}\right)\;,\\
  \vec{n}_\gamma&\equiv&
  \left(\vec{n}^{(1)}_\gamma,\vec{n}^{(2)}_\gamma,\ldots,\vec{n}^{(N_z)}_\gamma\right)\;,\\
  \mat{P}_{\gamma\kappa}&\equiv&
      {\rm
	diag}\{\hat{\mat{P}}_{\gamma\kappa},\hat{\mat{P}}_{\gamma\kappa}
      ,\ldots,\hat{\mat{P}}_{\gamma\kappa}\}\;.
\end{eqnarray}
The round brackets, grouping together the vector arguments, should be
understood as big vectors obtained by piling up all embraced vectors
on top of each other.  Employing the linear relation between shear and
convergence and plugging them into the Saskatoon filter,
Eq. \Ref{eq:shakatoon}, with \mbox{$\mat{R}=\mat{P}_{\gamma\kappa}$},
\mbox{$\mat{S}=\ave{\vec{\kappa}\vec{\kappa}^{\rm t}}^\prime$} and
\mbox{$\mat{N}=\ave{\vec{n}_\gamma\vec{n}_\gamma^\dagger}$} gives an
estimator, $\vec{\kappa}_{\rm MV}$, for the lensing convergence seen
by the sources within the various redshift bins.

This estimator is ideally, only for exactly vanishing B-modes in the
cosmic shear signal, a purely real vector. The imaginary part of
$\vec{\kappa}_{\rm MV}$ is at best pure noise, but contains in general
a significant signal owing to a possible B-mode signal in the cosmic
shear field. Thus, the imaginary part of $\vec{\kappa}_{\rm MV}$ can
be used as parallel B-mode reconstruction, whereas the real part is
the sought E-mode of the convergence tomography. E-mode and B-mode are
interchangeable by rotating the source ellipticities by $45^\circ$.

For the statistical uncertainty, needed for $\mat{N}$, of the source
ellipticity attached to the $k$th grid pixel and source bin $i$, we
take
\begin{equation}
  \Ave{\vec{n}^{(i)}_\gamma[\vec{n}^{(j)}_\gamma]^{\rm t}}_{kl}=
  \delta^{\rm K}_{kl}\delta^{\rm
  K}_{ij}\frac{[\sigma^{(i)}_\epsilon]^2}{N^{(i)}_k}\;,
\end{equation}
with
\begin{eqnarray}
  \left[\sigma^{(i)}_\epsilon\right]^2&\equiv&
  \left[\sum_lw^{(i)}_l\right]^{-1}\sum_lw^{(i)}_l|E^{(i)}_l|^2\;,\\
  N^{(i)}_k&\equiv&
  \left[\sum_lw^{(i)}_l\right]^{2}\left[\sum_l\left[w^{(i)}_l\right]^{2}\right]^{-1}\;.
\end{eqnarray}
Here the sum for the overall ellipticity variance,
$[\sigma^{(i)}_\epsilon]^2$, runs over all source ellipticities
belonging to the $i$th source bin, and the sum for the effective
number of sources inside the $k$th pixel, $N^{(i)}_k$, only over
the statistical weights of sources belonging to that pixel.  For
pixels containing no sources, we set
$\sigma^{(i)}_\epsilon=\sigma_\infty$ with $\sigma_\infty=10^3$. The
noise covariance is a diagonal matrix, $\delta^{\rm K}_{kl}$ and
$\delta^{\rm K}_{ij}$ denote Kronecker symbols, since cross-correlations of
the intrinsic ellipticities of sources are neglected.

Although it is assumed for the scope of this paper that intrinsic
ellipticities of sources are uncorrelated to any other relevant
quantity, it may be possible to accommodate also cases where this is
not true. For example, if we expect intrinsic alignments between
sources of the same or different z-bins, i.e. correlations of
intrinsic ellipticities
\citep{2000MNRAS.319..649H,2000ApJ...545..561C,2001ApJ...559..552C}.
This case can be treated by adding the two-point correlation of
intrinsic alignments, $\xi_{\rm ia}^{(ij)}(\theta)$ as additional
noise contaminant to $\mat{N}$:
\begin{equation}
  \Ave{\vec{n}^{(i)}_\gamma[\vec{n}^{(j)}_\gamma]^{\rm t}}_{kl}=
  \delta^{\rm K}_{kl}\delta^{\rm
  K}_{ij}\frac{[\sigma^{(i)}_\epsilon]^2}{N^{(i)}_k}+ \xi_{\rm
  ia}^{(ij)}(\theta_{kl})\;.
\end{equation}

Another possible known contaminant of lensing applications is the
possible correlation between the intrinsic alignment of sources and
the lensing signal \citep{2004PhRvD..70f3526H}. This would give rise
to a correlation between noise and signal,
$\ave{\mat{s}\mat{n}^\dagger}\ne\mat{0}$, which was excluded here from
the very beginning of the derivation of the linear filters and is
being assumed to be negligible for this work. We defer the treatment
of this contaminant to a future paper but already outline in Appendix
\ref{sect:shearintr} how this can in principle be included in the
estimator.

The signal covariance, $\mat{S}$, is the two-point correlation
function of the lensing convergence at the separation of two grid
points, which is
\begin{equation}
  \Ave{[\vec{\kappa}^{(i)}]_k[\vec{\kappa}^{(j)}]_l}^\prime=
  \Ave{\gamma^{(i)}(\vec{\theta}_k)[\gamma^{(j)}(\vec{\theta}_l)]^\ast}^\prime=
  \xi^{(ij)}_+(\theta_{kl})\;,
\end{equation}
where $\xi^{(ij)}_+$ is a two-point correlation function of the cosmic
shear \citep[e.g.][]{2001PhR...340..291B} between the $i$th and $j$th
source bin, $\gamma^{(i)}(\vec{\theta}_k)$ and
$\gamma^{(j)}(\vec{\theta}_l)$.
  
\subsection{Density contrast on lens planes}
\label{sect:densityest}

We can go one step further than the lensing convergence by
recognising that the convergence $\vec{\kappa}^{(i)}$ itself, as
seen in the sources of the $i$th redshift bin, is within the
framework of general relativity a linear transformation of the matter
density contrast, $\delta$, along the line-of-sight:
\begin{eqnarray}\label{kappa1}
  [\vec{\kappa}^{(i)}]_k&=&
  \frac{3H_0^2}{2c^2}\Omega_{\rm m}\times\\\nonumber
  &&
  \int_0^\infty\d w\frac{\overline{W}^{(i)}(w)f_{\rm K}(w)}{a(w)}
  \delta\left(f_{\rm K}(w)\vec{\theta}_k,w\right)\;,\\
   \overline{W}^{(i)}(w)&\equiv&
   \int_0^{w}\d w^\prime\,
   \frac{f_{\rm K}(w^\prime-w)}{f_{\rm K}(w^\prime)}
   \left[p_z^{(i)}(z)\frac{\d z}{\d w}\right]_{z=z(w^\prime)}\!\!\!\!\;.
\end{eqnarray}
The function $f_{\rm K}(w)$ is the angular diameter distance as
function of the comoving radial distance $w$ and the curvature,
$K$, of the fiducial cosmological model:
\begin{eqnarray}
  f_{\rm K}(w) &=& \left\{
  \begin{array}{ll}
    K^{-1/2}\sin(K^{1/2}w) & (K>0)\\
    w & (K=0)\\
    (-K)^{-1/2}\sinh[(-K)^{1/2}w] & (K<0) \\
  \end{array}\right.\;.
\end{eqnarray}
Furthermore, we denote the vacuum speed of light and the cosmological
scale factor by $c$ and $a(w)$, respectively.  For the density
contrast, we have chosen the comoving coordinate frame such that the
first argument $\vec{x}_\perp$ in $\delta(\vec{x}_\perp,x_\parallel)$
denotes a position in a plane perpendicular to the line-of-sight, and
the second argument, $x_\parallel$, the distance of the plane along
the line-of-sight. Note that we are assuming a flat sky.

\begin{figure}
  \begin{center}
    \psfig{file=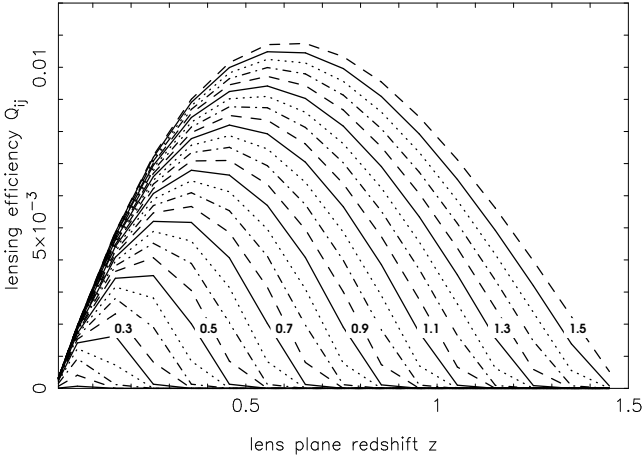,width=85mm,angle=0}
  \end{center}
  \caption{\label{fig:lenseff}Lensing efficiency, $Q_{ij}$, for
    \mbox{$j=1\ldots15$} lens-planes, equally spaced between redshifts
    \mbox{$z=0,1.5(\Delta z=0.1)$}, and \mbox{$i=1\ldots30$} source
    redshift bins, equally spaced between \mbox{$z=0,1.5(\Delta
      z=0.05)$}. The numbers labelling some of the lines denote the mean
    source redshift; the source redshift increases from left to right.
    The galaxy distribution within the redshift bins is assumed to be
    flat.}
\end{figure}

For a practical approximation of Eq. \Ref{kappa1}, we postulate
$\delta$ to be constant over $l=1\ldots N_{\rm lp}$ comoving distance
ranges $x_\parallel\in[w_l,w_{l+1}]$:
\begin{eqnarray}\label{deltadef}
  [\vec{\kappa}^{(i)}]_k&\approx&
  \sum_{l=1}^{N_{\rm lp}}
  Q_{il}\,\delta^{(l)}_k\;,\\
  Q_{il}&\equiv&
  \frac{3H_0^2}{2c^2}\Omega_{\rm m}
  \int_{w_l}^{w_{l+1}}\d w\frac{\overline{W}^{(i)}(w)f_{\rm
      K}(w)}{a(w)}\;.
\end{eqnarray}
This means, we assume that the 3-D density contrast stays constant
along the line-of-sight as long as we stay inside the same matter
slice; $\delta^{(i)}_k$ is here the average 3-D density contrast, and
not the total, over the width of the lens-slice integrated 3-D contrast
as can also be found in the literature.  The coefficients $Q_{il}$
(lensing efficiency) define the response of the convergence tomography
to the matter density on the individual lens-planes. A mean density
contrast $\bar{\delta}$, over the radial width of the $i$th slice,
induces on average shear polarisation of sources belonging to the
$j$th redshift bin of amplitude $Q_{ji}\bar{\delta}$. The coefficients
$Q_{ij}$ are independent of the line-of-sight direction,
Fig. \ref{fig:lenseff} displays an example for $Q_{il}$.

Due to the approximation in
Eq. \Ref{deltadef}, the mean matter density contrast within a slice
is:
\begin{eqnarray}
  \delta^{(i)}_k&=& \frac{\int_{w_i}^{w_{i+1}}\d
    w\,p^{(i)}(w)\delta\left(f_{\rm
      K}(w)\vec{\theta}_k,w\right)}{\int_{w_i}^{w_{i+1}}\d
    w\,p^{(i)}(w)}\;,\\ 
  p^{(i)}(w)&\equiv&\frac{\overline{W}^{(i)}(w)f_{\rm
      K}(w)}{a(w)}\;,
\end{eqnarray}
hence a mean weighed with $p^{(i)}(w)$.  For relatively narrow
lens-planes, the weights $p^{(i)}(w)$ will be roughly constant over
the slice width so that they can overall be approximated by top-hat
functions with width $\Delta w_i=w_{i+1}-w_i$. This is what we will
assume for the remainder of the paper.

As before, we can cast the relation between mean matter density and
convergence tomography into a compact notation by writing:
\begin{equation}
  \vec{\kappa}=\mat{Q}\,\vec{\delta}\;,
\end{equation}
where
\begin{eqnarray}
  \vec{\delta}&\equiv&\left(
  \vec{\delta}^{(1)},\vec{\delta}^{(2)},\ldots,\vec{\delta}^{(N_{\rm
      lp})}\right)\;,\\ \mat{Q}&\equiv&\left(
  \begin{array}{llll}
    \mat{1}Q_{11}&\mat{1}Q_{12}&\ldots&\mat{1}Q_{1N_{\rm lp}}\\
    \mat{1}Q_{21}&.&\ldots&\mat{1}Q_{2N_{\rm lp}}\\
    .&.&\ldots&.\\
    \mat{1}Q_{N_z1}&.&\ldots&\mat{1}Q_{N_zN_{\rm lp}}
  \end{array}\right)\;,
\end{eqnarray}
and where $\mat{1}$ is the $N_{\rm g}\times N_{\rm g}$ unity
matrix. The relation between the observable source ellipticities and
the lens-plane density contrast is therefore:
\begin{equation}\label{gammadeltaprojection}
  \vec{\gamma}=\mat{P}_{\gamma\kappa}\mat{Q}\vec{\delta}+\vec{n}_\gamma\;.
\end{equation}
As before for the lensing convergence tomography, an estimator for the
average density contrast, $\vec{\delta}_{\rm MV}$, on the lens-planes
is given by Eq. \Ref{eq:shakatoon} with
\mbox{$\mat{R}=\mat{P}_{\gamma\kappa}\mat{Q}$},
\mbox{$\mat{N}=\ave{\vec{n}_\gamma\vec{n}_\gamma^\dagger}$} and
\mbox{$\mat{S}=\mat{S}_\delta\equiv\ave{\vec{\delta}\vec{\delta}^{\rm
t}}^\prime$}.  For an efficient, practical implementation of the
filter see Appendix \ref{sect:kdtree}.

Analogous to the estimator $\vec{\kappa}_{\rm MV}$, $\vec{\delta}_{\rm
MV}$ will be a complex vector although the density contrast is by
definition a real number. The real part of $\vec{\delta}_{\rm MV}$
will be the mass reconstruction due to the cosmic shear E-mode,
whereas the imaginary part has to be attributed to the cosmic shear
B-mode.

\begin{figure}
  \begin{center}
    \psfig{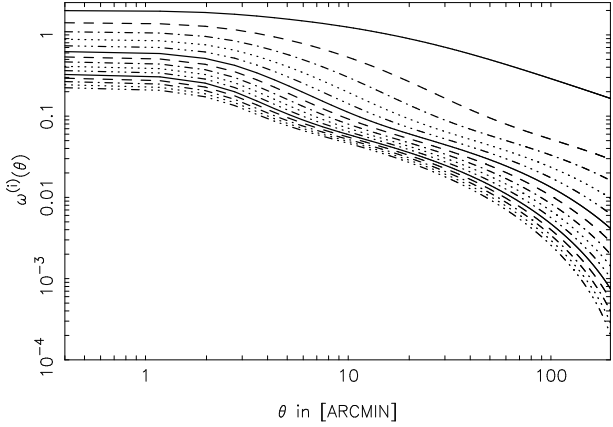}
  \end{center}
  \caption{\label{fig:corrplot}Expected projected two-point
    correlation, \mbox{$\omega^{(i)}(\theta)$}, of the density
    contrast situated inside \mbox{$i=1\ldots9$} equally spaced,
    \mbox{$\Delta z=0.1$}, lens-planes with redshift limits between
    \mbox{$z=0\ldots0.9$}. The mean redshift of the lens-planes
    decreases from the bottom to the top line. The correlation
    functions have been smoothed for a grid with pixel scale
    $1^\prime$.}
\end{figure}

\subsubsection{Transverse Wiener filter}
 
How can a prior on the signal covariance $\mat{S}_\delta$ be computed
for a given fiducial cosmological model? By exploiting the Limber
equation in Fourier space \citep{2001PhR...340..291B}, the relation
between 3-D matter density and projected density can be used to work
out the second-order correlations of $\delta^{(i)}_k$:
\begin{eqnarray}\label{deltasmooth}
  \Ave{\delta^{(i)}_k\delta^{(j)}_l}^\prime&=&
  \delta^{\rm K}_{ij}\,\omega^{(i)}(|\vec{\theta}_{kl}|)\;,\\
  \label{deltasmooth2}
  \omega^{(i)}(\theta)&=&\frac{1}{2\pi} \int_0^\infty\d\ell\ell\,
  P^{(i)}_\delta(\ell)\,|F(\Theta_{\rm s}\ell)|^2\,J_0(\ell\theta)\;,\\\label{deltasmooth3}
  P^{(i)}_\delta(\ell)&=& \frac{1}{(\Delta w_i)^2} \int_{w_i}^{w_{i+1}}
  \!\!\!\frac{\d w}{[f_{\rm K}(w)]^2}P_{\rm 3d}\left(\frac{\ell}{f_{\rm
      k}(w)},w\right)\;.
\end{eqnarray}
Here $P_{\rm 3d}(k,w)$ is the fluctuation power spectrum of the
3-D matter distribution for a spatial scale $k$ at a comoving distance
$w$. For the scope of this paper, the approximation of
\citet{2003MNRAS.341.1311S} will be used. The function
$P_\delta^{(i)}(\ell)$ quantifies the fluctuation power spectrum of
$\delta^{(i)}(\vec{\theta})$, the projected density contrast on the
lens-planes. Note that cross-correlations between $\delta^{(i)}$'s
belonging to different lens-planes vanish in the Limber
approximation. Fig. \ref{fig:corrplot} gives an example for
$\omega^{(i)}(\theta)$. The window kernel $F(\ell)$ accounts for the
binning on the grid, it is the Fourier transform of a grid pixel,
which we assume as an approximation to be a circular top-hat function,
$A=\pi\Theta_{\rm s}^2$, with radius $\Theta_{\rm s}$, hence:
\begin{equation}\label{eq:window}
  F(\Theta_{\rm s}\ell)=\frac{2J_1(\Theta_{\rm s}\ell)}{\Theta_{\rm s}\ell}\;.
\end{equation}
$J_n(x)$ is the Bessel function of first kind.

We call this prior on the correlations of density contrasts in
different directions a \emph{transverse} Wiener filter.

\subsubsection{Radial Wiener filter}
\label{sect:radialfilter}

\citet{2002PhRvD..66f3506H} discuss another Wiener filter, also
applied in \citet{2003MNRAS.344.1307B} for the 3-D potential, that puts
priors only on the correlation between structures (average density
contrast inside pixel needle light cones) in the radial direction,
i.e.
\begin{equation}\label{eq:hukeeton}
  \Ave{\delta_k^{(i)}\delta_l^{(j)}}^\prime= \delta_{kl}
  \int\frac{\d^3\vec{k}}{(2\pi)^3}
  \hat{F}_i(\vec{k})\hat{F}^\ast_j(\vec{k})P^{(ij)}_\delta(k)
  \end{equation}
where
\begin{eqnarray}
  \hat{F}_i(\vec{k})&\equiv&{\rm e}^{{\rm i}k_\|
    \bar{w}_i}\frac{\sin{(k_\|\Delta w_i/2)}}{k_\|\Delta
    w_i/2}F(k_\perp \bar{w}_i\Theta_{\rm
    s})\;,\\
      P_\delta^{(ij)}(k)&\equiv&\sqrt{P_\delta(k,\bar{w}_i)P_\delta(k,\bar{w}_j)}\;,\\
      k^2&\equiv&k_\|^2+k_\perp^2\;.
\end{eqnarray}
The solid angle $\Theta_{\rm s}$ denotes the radius of a circular
pixel in radians and $\bar{w}_i=w_i+\Delta w_i/2$ is the comoving
(radial) distance of the centre of the $i$th lens-plane.

In the total absence of correlations between pixels of different
line-of-sights, the angular power spectra are flat power spectra
(white spectra), which for a circular top-hat smoothing window
equals\footnote{As a related example, take the intrinsic shear
ellipticity noise which has the flat power
$\sigma^2_\epsilon\bar{n}^{-1}$. The intrinsic noise inside a circular
pixel is then $\sigma^2_\epsilon/\bar{n}\pi\Theta_{\rm
s}^2=\sigma^2_\epsilon\bar{N}^{-1}$, where $\bar{N}$ is the mean
number of sources inside a pixel. Hence, we would expect the intrinsic
noise variance to go down as $\propto1/\sqrt{\bar{N}}$ (Poisson shot
noise).}
\begin{equation}
  P_{ij}\equiv
  \pi\Theta_{\rm s}^2\ave{\delta_l^{(i)}\delta_l^{(j)}}^\prime\;.
\end{equation}
We use this prior for the \emph{radial} Wiener filter. Obviously, the
amplitude of the flat signal power depends on the radius, $\Theta_{\rm
s}$, of the pixels since Eq. \Ref{eq:hukeeton} is also smoothing
density fluctuations in transverse direction.  The following
discussion will adopt $\Theta_{\rm s}=1^\prime$. This may appear too
specific, but a change of the radius will only rescale the overall
amplitude of $P_{ij}$ which, in turn, will solely act as a rescaling
of the tuning parameter $\alpha$. The essential feature of the radial
filter, however, namely its scale-independence, will remain unaffected
by the choice of $\Theta_{\rm s}$.

We find that $P_{ij}$ is close to diagonal with negative off-diagonal
elements roughly two orders of magnitude or more smaller than the
diagonal elements. Thus, the main difference between our radial and
transverse Wiener filter is the scale-independence of the radial
filter, the off-diagonal elements are not of much importance which we
checked by setting them to zero.  

\changed{The reader may be reminded that the two different Wiener
priors discussed here -- the radial and the transverse filter --
differ only in the way assumptions about correlations between pixels
on the lens planes are made. The terms do not mean that radial and
transverse modes are reconstructed separately. All modes are
determined simultaneously during the process of reconstruction, no
decomposition into radial and transverse modes is done.}

\subsection{Fourier space reconstructions}
\label{sect:densitycontrast}
For a discussion of the noise properties of the 3-D mass reconstruction
in the following section a Fourier representation of the problem is
extremely helpful. Moreover, it allows with some idealistic
assumptions to perform a quick-and-dirty reconstruction in Fourier
space which is of practical interest, see Appendix
\ref{sect:fourierrecon}.

An idealised survey is considered. Such a survey has either periodic
boundary conditions or is infinite in extent (flat sky), and the
statistical noise $\vec{n}^{(i)}_\gamma$ is homogeneous for all source
redshift bins. The latter means that the statistical uncertainty in
the shear estimate is equal for each grid cell of the same redshift
bin.

In this case the matter density contrast, and all other fields, can
equivalently be expressed in terms of its angular modes
\begin{eqnarray}
  \vec{\delta}(\vec{\theta})&=&\int_{{\rm R}^2}\frac{\d^2\vec{\ell}}{(2\pi)^2}\,
  \tilde{\vec{\delta}}(\vec{\ell})\,{\rm e}^{+{\rm
      i}/2(\vec{\theta}\vec{\ell}^\ast+\vec{\theta}^\ast\vec{\ell})}\;,\\
  \vec{\tilde{\delta}}(\vec{\ell})&=&\int_{\rm R^2}\d^2\vec{\theta}\,
  \vec{\delta}(\vec{\theta})\,{\rm e}^{-{\rm
      i}/2(\vec{\theta}\vec{\ell}^\ast+\vec{\theta}^\ast\vec{\ell})}\;,
\end{eqnarray}
which defines the density contrast on a typical scale
\mbox{$2\pi/|\vec{\ell}|$}. The advantage is that now only angular
modes of the same $\vec{\ell}$ couple to produce the angular modes of
the shear tomography because the conversion from lensing convergence
to shear is a convolution,
\begin{equation}
  \tilde{\gamma}^{(i)}(\vec{\ell})=
  D(\vec{\ell})\sum_{j=1}^{N_{\rm
      lp}}Q_{ij}\tilde{\delta}^{(j)}(\vec{\ell})+
  \tilde{n}^{(i)}_\gamma(\vec{\ell})\;,
\end{equation}
where \mbox{$D(\vec{\ell})\equiv\vec{\ell}/\vec{\ell}^\ast$}
\citep{1993ApJ...404..441K} and \mbox{$\vec{\ell}=\ell_1+{\rm
    i}\ell_2$} with $\ell_1$ and $\ell_2$ being the Cartesian components
of the angular mode.

Devising the filter Eq. \Ref{eq:shakatoon} of Sect. \ref{sect:mlinv}
one gets for the MV-density mode
\begin{equation}\label{estimatorfourier}
  \tilde{\vec{\delta}}_{\rm MV}(\vec{\ell})= D^\ast(\vec{\ell})
  \left[\alpha\tilde{\mat{S}}_\delta^{-1}+\mat{Q}^{\rm t}
    \tilde{\mat{N}}^{-1}\mat{Q}\right]^{-1}
  \mat{Q}^{\rm t}\tilde{\mat{N}}^{-1}
  \tilde{\vec{\gamma}}\;.
\end{equation}
Note that $\tilde{\vec{\delta}}$ and $\tilde{\vec{\gamma}}$ (the
Fourier transform of the gridded shear catalogues) are written here
in a compact vectorial form with mere $N_{\rm lp}$/$N_{\rm z}$
elements, grouping all lens-planes/source bins together, as outlined
in the foregoing sections; $\alpha$ is the usual tuning parameter.
Here we furthermore use
\begin{equation}
  [\tilde{\mat{N}}]_{ij}\equiv\delta^{\rm K}_{ij}\frac{[\sigma_\epsilon^{(i)}]^2}{\bar{n}_i}\;,
\end{equation} 
which -- for the sake of a simplistic noise model used in the next
section -- assumes a shot-noise term only\footnote{In the presence of
correlations between intrinsic source ellipticities, the noise
covariance needs additionally to be offset by
\mbox{$[\tilde{\mat{N}}]_{ij}\mapsto
[\tilde{\mat{N}}]_{ij}+P^{(ij)}_{\rm ia}(\ell)$}, where $P^{(ij)}_{\rm
ia}(\ell)$ is the power spectrum of the intrinsic correlations between
the $i$th and $j$th source bin as function of scale $|\vec{\ell}|$.};
$\bar{n}_i$ is the mean number density of sources belonging to the
$i$th redshift bin.

The \emph{transverse Wiener filter} constraints the expected signal
power by
\begin{equation}
  [\tilde{\mat{S}}_\delta]_{ij}\equiv
  \delta^{\rm K}_{ij}P^{(i)}_\delta(|\vec{\ell}|)\;.
\end{equation}
Noise and signal are subject to the same smoothing if we bin our data
on a grid. Therefore, the effect of a smoothing window cancels out in
the filter, since $\tilde{\mat{S}}_\delta$ and
$\tilde{\mat{N}}$ appear always as ratios in the filter.

The (expected) signal $\tilde{\mat{S}}_\delta$ is a diagonal matrix
for transverse filtering in the Fourier space representation. This may
change, if we choose to apply \emph{radial Wiener filtering}, where
only correlations between matter densities along the same
line-of-sight are regularised. Since correlations between different
directions are neglected, the power spectra used for regularisation
are flat power spectra (Sect. \ref{sect:radialfilter}), independent of
$\ell$:
\begin{equation}
  [\tilde{\mat{S}}_\delta]_{ij}=P_{ij}\;.
\end{equation}

\subsection{Gravitational potential on lens planes}
\label{sect:gravpot}

We may ask the question how the MV-estimator of $\vec{\delta}$ is
related to the MV-estimator of 
\begin{equation}
  \vec{\phi}\equiv\mat{F}\vec{\delta}
\end{equation}
that is thought to be a linear, \emph{invertible} transformation of
$\vec{\delta}$. 

Such cases could be (lens-plane-) smoothed density reconstructions,
for which $\mat{F}^{(i)}$ would be a (invertible) smoothing operator
acting on $\vec{\delta}^{(i)}$, or the 2-D gravitational potential on
the lens-planes which can be pictured as a special type of smoothing
of $\vec{\delta}^{(i)}$ \citep{2003MNRAS.344.1307B}:
\begin{equation}\label{eq:gravpot}
  \phi(\vec{\theta})=\frac{1}{\pi}\int_\Omega\d^2\vec{\theta}^\prime
  \delta(\vec{\theta}^\prime)\ln{|\vec{\theta}^\prime-\vec{\theta}|}\;,
\end{equation}
or in a form for a discrete grid assuming that the grid sizes for
$\vec{\phi}$ and $\vec{\delta}$ are equal:
\begin{equation}
  [\mat{F}^{(i)}]_{kl}=
  \frac{A}{\pi}\ln{|\vec{\theta}_l-\vec{\theta}_k|}~\;;~
  \vec{\phi}^{(i)}=\mat{F}^{(i)}\vec{\delta}^{(i)}\;.
\end{equation} 
The density contrast is constant over the size of one (round)
pixel. For \mbox{$\vec{\theta}_l=\vec{\theta}_k$} one finds therefore
\mbox{$[\mat{F}^{(i)}]_{kk}=-A/2\pi$}.  Note that Eq. \Ref{eq:gravpot}
gives only one possible solution for the 2-D gravitational potential
because the potential for a given $\delta^{(i)}(\theta)$ is not unique.

The new, smoothed $\vec{\phi}$, is related to the data, $\vec{d}$,
by
\begin{eqnarray}
  \vec{d}&=&\mat{R}\mat{F}^{-1}\vec{\phi}+\vec{n}\;,\\
  \mat{F}&\equiv&{\rm diag}\left\{\mat{F}^{(1)},\mat{F}^{(2)},\ldots,\mat{F}^{(N_{\rm
      lp})}\right\}\;.
\end{eqnarray}
Following the arguments in Sect. \ref{sect:mlinv} it is easy to show
that the MV-estimator for $\vec{\phi}$ is
\begin{equation}
  \vec{\phi}_{\rm MV}=\mat{F}\vec{\delta}_{\rm MV}.
\end{equation}
This means, once we have got the MV-solution of $\vec{\delta}$ we can
simply apply $\mat{F}$ to it in order to acquire the MV-solution of
$\vec{\phi}$. We do not have to go through the process of another
full MV-reconstruction, if we want to further smooth a
MV-solution. That is, provided the smoothing is invertible. Moreover,
the covariance of $\vec{\phi}_{\rm MV}$ is
\begin{equation}
  {\rm Cov}{(\vec{\phi}_{\rm MV})}= \mat{F}^\dagger{\rm
    Cov}{(\vec{\delta}_{\rm MV})}\mat{F}\;.
\end{equation}

\section{Bias and noise properties of reconstructions}
\label{sect:fourier}

As shown in \citet{2002PhRvD..66f3506H} a mass reconstruction, even
with as many sources as $100~\rm arcmin^{-2}$ and with heavy
smoothing, is very noisy. Therefore, an additional regularisation of
the reconstruction, by means of Wiener filtering for instance, is an
absolute necessity. With regularisation, however, a careful analysis
of the remaining statistical uncertainties and biases in the
reconstruction will be required. This will be the focus of this
section.

\subsection{Average signal-to-noise of reconstructions}
\label{sect:averagesn}

Devising the MV-filter in Fourier space, Eq. \Ref{estimatorfourier},
offers a convenient way to estimate the covariance of the noise in the
reconstruction. The following conclusions, however, will be based on
the simplifying assumption that the noise pattern is homogeneous and
that there are no gaps in the data, in particular the shear pattern of
the full (infinite) flat sky is known.

We start by considering the signal-to-noise for individual
$\ell$-modes and assuming that the matter fluctuations of the
large-scale structure that provide the signal in the shear catalogue
have cosmic average amplitudes.

The covariance of the statistical uncertainties in
Eq. \Ref{estimatorfourier} is according to Eq. \Ref{covariance}:
\begin{equation}\label{fouriercovdelta}
  \tilde{\mat{N}}_{\rm MV}(\alpha)
  =\widetilde{\mat{W}}\tilde{\mat{X}}\widetilde{\mat{W}}^\dagger	 
\end{equation}
with
\begin{equation}
  \widetilde{\mat{W}}\equiv
  \tilde{\mat{S}}_\delta[\alpha\tilde{\mat{X}}+\tilde{\mat{S}}_\delta]^{-1}~{\rm and}~\tilde{\mat{X}}\equiv[\mat{Q}^{\rm t}
    \tilde{\mat{N}}^{-1}\mat{Q}]^{-1}\;.
\end{equation}
The latter is the noise covariance without a Wiener prior or if the
prior is unimportant.

How much signal-to-noise can we expect on average for a 3-D mass
reconstruction as function of angular scale? To answer this question,
we compare, for a $\Lambda\rm CDM$ fiducial cosmological model, the
expected signal power as function of scale $\ell$, $\tilde{\mat{S}}$,
getting through the filter (Eq. \ref{scovariance}),
\begin{equation}\label{fourierpowdelta}
  \tilde{\mat{S}}_{\rm MV}(\alpha)
  =\widetilde{\mat{W}}\tilde{\mat{S}}\widetilde{\mat{W}}^\dagger\;,
\end{equation}
to the noise power:
\begin{equation}
  \frac{\rm S}{\rm N}(\alpha)~{\rm of~}\tilde{\delta}^{(i)}_{\rm
    MV}(\vec{\ell})\approx 
  \sqrt{\frac{[\widetilde{\mat{W}}\tilde{\mat{S}}\widetilde{\mat{W}}^\dagger]_{ii}}
    {[\widetilde{\mat{W}}\tilde{\mat{X}}\widetilde{\mat{W}}^\dagger]_{ii}}}\;.
\end{equation}
As expected density fluctuation power we take the prediction based
upon the Limber equation.

Since the noise covariance, $\tilde{\mat{N}}$, scales as
$\propto\sigma_\epsilon^2/\bar{n}$, we can derive useful scaling
relations for the signal and noise power with a fixed redshift
distribution of sources. One finds that the tuning parameter inside
the filter scales as
\begin{equation}
  \alpha\mapsto
  \alpha\times
  \left(\frac{\sigma^\prime_\epsilon}{\sigma_\epsilon}\right)^2
  \left(\frac{\bar{n}_{}}{\bar{n}_{}^\prime}\right)=\alpha^\prime\;,
\end{equation}
if we change the intrinsic shape variance from $\sigma_\epsilon$ to
$\sigma^\prime_\epsilon$ or the source number density from $\bar{n}$
to $\bar{n}^\prime$. Thus, increasing the number density to
$\bar{n}^\prime$, for instance, yields a filter corresponding to
$\bar{n}$ but with a different, smaller $\alpha$. Similarly, we get:
\begin{eqnarray}
  \tilde{\mat{N}}_{\rm MV}(\alpha)&\mapsto&
  \left(\frac{\sigma^\prime_\epsilon}{\sigma_\epsilon}\right)^2
  \left(\frac{\bar{n}_{}}{\bar{n}_{}^\prime}\right)\times
  \tilde{\mat{N}}_{\rm MV}(\alpha^\prime)\;,\\
  \tilde{\mat{S}}_{\rm MV}(\alpha)&\mapsto&
  \tilde{\mat{S}}_{\rm MV}(\alpha^\prime)\;,
\end{eqnarray}
so that the signal-to-noise of modes scales as
\begin{equation}
  {\rm\frac{S}{N}(\alpha)}
  \mapsto
  \left(\frac{\sigma_\epsilon}{\sigma^\prime_\epsilon}\right)
  \sqrt{\frac{\bar{n}^\prime_{}}{\bar{n}_{}}}
  \times{\rm\frac{S}{N}(\alpha^\prime)}\;.
\end{equation}
This property is useful for preparing plots of the estimated
signal-to-noise spanning a wide range of fiducial surveys.

\begin{figure*}
  \begin{center}
    \psfig{file=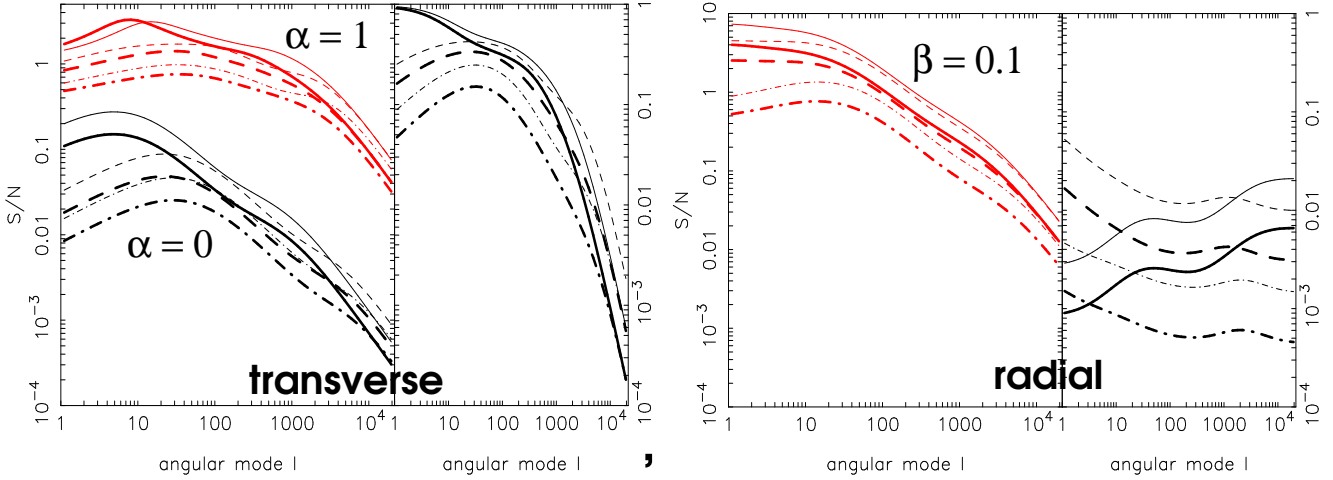,width=175mm,angle=0}
  \end{center}
  \caption{\label{fig:snpower} Signal-to-noise ratio of reconstructed
    matter density modes as function of angular scale for three
    different lens-planes (solid line: $\bar{z}=0.16$, dashed:
    $\bar{z}=0.76$, dashed-dotted: $\bar{z}=1.05$) using transverse
    (left figure) or radial (right figure) Wiener filtering (top
    lines; $\beta=0.1$ for the radial filter, $\alpha=1.0$ otherwise)
    and no prior (bottom lines; $\alpha=0$). The tuning parameters
    of the radial, $\beta$, and the transverse filter, $\alpha$,
    are not equivalent (see text). The reconstruction uses $15$
    equally spaced lens-planes, between \mbox{$z=0\ldots1.5$}, and
    $30$ equally space source redshift bins, between
    \mbox{$z=0\ldots1.5$}, with a source number density of $30\,{\rm
    arcmin}^{-2}$ (thin lines: $100\,{\rm arcmin}^{-2}$) and
    $\sigma_\epsilon=0.3$. The mean redshift of the sources is
    \mbox{$\bar{z}=0.85$}. The right sub-panels show the attenuation
    factor $\sqrt{[\tilde{\mat{S}}_{\rm
    MV}(\alpha)]_{ii}/P_\delta^{(i)}(\ell)}$ ($=1$ for no damping)
    inflicted by the Wiener filter as function of scale. Plots for
    $\alpha=0$ are identical for both the radial and the transverse
    filter and are therefore only shown in the left panel.}
\end{figure*}

Fig. \ref{fig:snpower} is a particular example of the signal-to-noise
of reconstructed density modes using realistic survey parameters. For
the source distribution with redshift a function
\begin{equation}\label{eq:fiducialpz}
  p(z)\propto z^2{\rm e}^{-\left(\frac{z}{z_0}\right)^{1.5}}
\end{equation}
with \mbox{$z_0=0.57$} was used; the mean redshift is
$\bar{z}=0.85$. The distribution was truncated beyond a redshift of
\mbox{$z=1.5$}. The redshift distribution is somewhat shallow but not
too unrealistic for contemporary surveys if we take into account that
source redshifts are required.

Obviously, even for a very optimistic source density of $100\,\rm
arcmin^{-2}$, Wiener filtering is utterly necessary to improve the
signal-to-noise in the reconstructions.  Simple transverse smoothing
of the reconstruction obtained without Wiener prior, i.e. suppressing
high-$\ell$ modes, does not suffice as even the lowest $\ell$-modes
have a signal-to-noise less than unity.

The transverse or radial Wiener filter bring the signal up to a
signal-to-noise of roughly unity, somewhat higher on very large
angular scales. If we tune down the Wiener filter by lowering the
$\alpha$-parameter, we get a signal-to-noise that lies somewhere in
between of the full Wiener filter and the very noisy reconstruction
with no prior (not shown). These figures also demonstrate the
attenuation of low-S/N modes by the Wiener filter, i.e. the ratio
$\sqrt{[\tilde{\mat{S}}_{\rm MV}(\alpha)]_{ii}/P_\delta^{(i)}(\ell)}$.
This tells us that the Wiener filter will, on average, not recover the
original amplitude of a fluctuation mode. Instead, the amplitude will,
depending on the redshift of the lens plane and the scale $\ell$, be
attenuated. This is an unavoidable side-effect of Wiener filtering,
which, on the other hand, thereby improves the signal-to-noise in the
reconstruction.  Another side-effect concerning the radial
distribution of densities will be discussed in the next section.

For the transverse filter especially, small scale modes are damped,
which thereby acts as a low-pass filter performing an automatic
smoothing on the lens-planes. In case of a radial Wiener filter, the
damping is roughly the same on all scales due to the
scale-independence of the regularisation. 

Notice that in Fig. \ref{fig:snpower} the tuning parameters for the
transverse, $\alpha$, and radial filter, $\beta$, are not completely
equivalent, although they have the same effect: For our transverse
filter, the theoretical signal power is identical to the expected
signal power plugged into the filter as
$\tilde{\mat{S}}_\delta=\tilde{\mat{S}}$ (both based on Limber's
equation), whereas for the radial Wiener filter, the expected power
$\tilde{\mat{S}}$ -- still the same as before to have a comparison
assuming the same input signal -- is different from
$\tilde{\mat{S}}_\delta$ used inside the Wiener filter (flat power
spectrum).

\begin{figure}
  \begin{center}
    \psfig{file=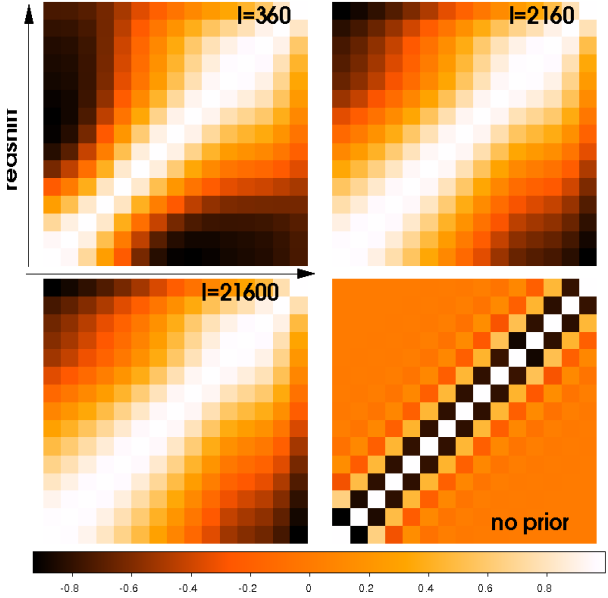,width=80mm,height=80mm,angle=0}
  \end{center}
  \caption{\label{fig:corr} Correlation of statistical errors of
    $\tilde{\delta}^{(i)}_{\rm MV}(\vec{\ell})$ and
    $\tilde{\delta}^{(j)}_{\rm MV}(\vec{\ell})$ for three distinct
    $\ell$ (transverse Wiener filter) and no prior or $\alpha=0$
    (bottom right), which is independent of $\ell$. This plot assumes
    cosmic average matter fluctuations. The Wiener filter assumes
    $\alpha=1$ and a galaxy number density of $30\,\rm arcmin^{-2}$.}
\end{figure}

The improvement of the signal-to-noise due to Wiener filtering comes
with another price that has to be paid. The statistical errors between
the modes of different lens-planes, which are weakly correlated if no
Wiener filtering is applied, become correlated. Fig. \Ref{fig:corr}
shows the matrix
 \begin{equation}
   r_{ij}(\ell)=
   \frac{[\tilde{\mat{N}}_{\rm MV}(\alpha)]_{ij}}
   {\sqrt{[\tilde{\mat{N}}_{\rm MV}(\alpha)]_{ii}[\tilde{\mat{N}}_{\rm MV}(\alpha)]_{jj}}}
 \end{equation}
for the previous survey parameters with \mbox{$30\,\rm arcmin^{-2}$}
source density. With no prior at all one finds that the errors between
different lens-planes quickly decorrelate in a oscillatory manner
(bottom right). As pointed out in \citet{2002PhRvD..66f3506H} this
particular form of error correlations is owed to the finite difference
approximation of a fourth-derivative in radial direction.

By means of the Wiener filter, however, the errors become correlated
over a wider range (the filter also smoothes in radial
direction). Errors between different $\ell$'s remain uncorrelated as
they were. For a signal-to-noise close to one, the error correlations
limit our ability to pin down the redshift of structures along the
line-of-sight, that is the redshift resolution of the
reconstruction. The correlations have also a positive side: the radial
oscillations become stretched out by the Wiener filter.

The foregoing way of estimating the signal-to-noise in a
reconstruction takes a quite pessimistic view. Namely, it presumes
that the matter fluctuations of the structure we would like to recover
are actually of an amplitude we expect for a cosmic average, i.e. we
expect $|\tilde{\delta}(\ell)|^2\sim P_\delta(\ell)$.  In fact, in
individual regions of space we will find matter concentration
exceeding the average (large galaxy clusters), or we will find less
matter (voids).

Let's quickly make an rough estimate of how much power above the
average we may can expect.  For a Gaussian random field -- probably
being a fair approximation of the mass density field on large scales
and, thus, the regime we can realistically hope to reconstruct -- the
expected variance of the fluctuation power is (exploiting Wick's
theorem):
\begin{equation}
  \Ave{P(\ell)^2}^\prime-\Ave{P(\ell)}^{\prime2}=2P(\ell)\;.
\end{equation}
Therefore, we may expect, owing to the natural variance in a Gaussian
random field, to find regions where the signal power is actually up to
four ($2\sigma$) times the average fluctuation power, enhancing the
signal-to-noise in Fig. \ref{fig:snpower} by a factor of two -- or
even more in the non-linear regime.  In Sect. \ref{sect:snhaloes} we
will work out the signal-to-noise of specific matter haloes.

\subsection{Response in Fourier space}
\label{sect:response}

\begin{figure}
  \begin{center}
    \psfig{file=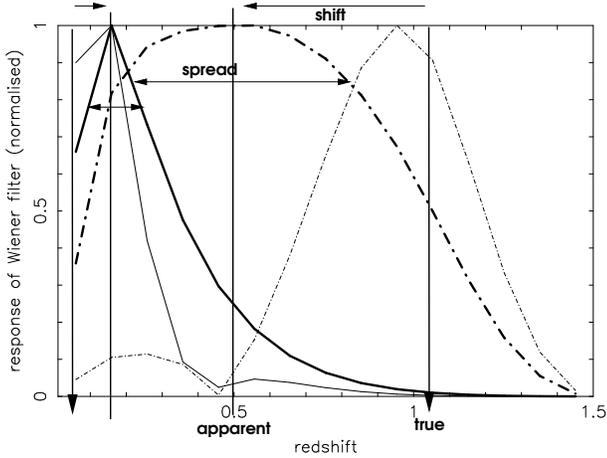,width=80mm,angle=0}
  \end{center}
  \caption{\label{fig:response} Examples for the response of the
    Wiener filtered (transverse) matter reconstruction algorithm to
    peaks in angular Fourier space of coherence scale $\ell=800$ in
    the true matter distribution at low, $z\sim0.05$, and high,
    $z\sim1.05$, redshift (downward arrows). The fiducial survey has
    the parameters as outlined in the caption of
    Fig. \ref{fig:lenseff} assuming a source density of $30\,\rm
    arcmin^{-2}$ with $\alpha=1$ (thick lines). In the
    3-D reconstruction, the localised peaks are shifted in redshift
    and smeared out, independent of the underlying signal, which
    limits our ability to reconstruct the matter distribution in
    radial direction. The thin lines correspond to a Wiener filter
    with $\bar{n}=100\,\rm arcmin^{-2}$ and $\alpha=1$. Increasing
    $\bar{n}$ or decreasing $\alpha$ reduces the shift and spread of
    the response.}
\end{figure}

There will be a fundamental limit in recovering the radial matter
distribution, even if we have peaks with a high signal-to-noise. As
discussed in Sect. \ref{sect:mlinv}, a Wiener based reconstruction is
unavoidably biased, unless the signal-to-noise of the data is
infinite. This bias is quantified by the bias matrix
\mbox{$\widetilde{\mat{W}}$} that relates the original vector of
$\ell$-mode coefficients to the (statistical average) vector of modes
in the reconstruction. It is a convolution applied to the true matter
distribution.

Specifically, the $i$th column of $\widetilde{\mat{W}}$ expresses how
the Wiener reconstruction, on average, responds in the reconstructed
density contrast field to a value $\tilde{\delta}^{(i)}(\vec{\ell})$
in the true matter distribution. Ideally, we would like to have a
unity bias matrix (unbiased estimator), but we find that, in the
reconstruction, the Fourier coefficient becomes rescaled
(attenuation), shifted and spread out in the radial direction,
contributing to other $\tilde{\delta}^{(j)}_{\rm MV}(\vec{\ell})$ with
$i\ne j$. This response for a certain $\ell$-mode is demonstrated by
Fig. \ref{fig:response}. In this particular but realistic example, an
angular mode of a lens-plane at redshift $z=1.05$ ($z=0.05$) is
shifted to $z\sim0.5$ ($z=0.15$) and smeared out by
$\sigma_z\approx0.3$ ($1\sigma$-width; $\sigma_z\approx0.1$) to
adjacent lens planes. The smearing of the peak is the reason for the
strong correlation of modes of neighbouring lens-planes, which has
been discussed in the forgoing section.

Especially structures at the upper redshift end of the survey a prone
to the shift and smearing. Both effects are solely a function of
angular scale, $\ell$, larger coherent structures (smaller $\ell$'s)
are easier to resolve than smaller structures. For small enough
$\ell$'s all redshifts are eventually shifted to one redshift and
similar response functions. 

\subsection{Point spread function}
\label{sect:psf}

The $\ell$-dependence of the response in the reconstruction as
described above is hard to relate to the question how well we can
recover the redshift matter clump as function of mass and
redshift. Therefore, the following translates the Fourier picture of
the reconstructions back to real (angular) space, where we further
study the Wiener filter impact.

Assume we would like to calculate the effect of the Wiener filter on a
halo sitting at some redshift, or equivalently on some lens plane $i$,
with projected density contrast profile
\begin{equation}
 \delta_{\rm 2d}(\vec{\theta})= \frac{1}{\Delta w}\int_0^\infty\d
 w\,\delta_{\rm 3d}(f_{\rm K}(w)\vec{\theta},w)\;.
\end{equation}
Furthermore, we consider for simplicity only profiles that are
rotationally invariant, i.e.
\begin{equation}
  \delta_{\rm 2d}(\vec{\theta})\equiv\delta(|\vec{\theta}|)\;.
\end{equation}
A single ``point'' would be just a Dirac delta function in direction of
$\vec{\theta}$.

The prefactor $\Delta w$ denotes the range of comoving distances
spanned by the matter slice the halo is located in.  In our framework,
$\delta_{\rm 2d}(\vec{\theta})$ is the average 3-D density contrast,
constant inside the slice represented by the lens-plane, and
\emph{not} the integrated 3-D density contrast over the width of a
slice, see Sect. \ref{sect:densityest}.

If we place the halo at the centre of our coordinate system,
$\vec{\theta}=\vec{0}$, and smooth the overdensity over the solid
angle, $A=\pi \Theta_{\rm s}^2$, of the circular pixel with radius
$\Theta_{\rm s}$ (our grid binning) centred on $\vec{\theta}=\vec{0}$,
\begin{equation}\label{eq:smooth}
  \bar{\delta}(\Theta_{\rm s})\equiv \frac{1}{A}\int_A\d^2\vec{\Theta}\,
  \delta_{\rm 2d}(\vec{\Theta})\;,
\end{equation}
we would find on average after Wiener filtering, performing the
integration over all modes $\ell$, a smoothed over-density of
\begin{eqnarray}\label{eq:pixelsignal0}
  \vec{\bar{\delta}}(\Theta_{\rm s},\alpha)&=&
  \frac{1}{(2\pi)^2}\int_{R^2}\d^2\vec{\ell}\,F(\ell
  \Theta_{\rm s})\,\widetilde{\mat{W}}
  \vec{\tilde{\delta}}(\ell)\\\label{eq:pixelsignalcirc} 
  &=&\frac{1}{2\pi}\int_0^\infty\d\ell\,\ell\,F(\ell \Theta_{\rm s})\,\widetilde{\mat{W}}\vec{\tilde{\delta}}(\ell)\;,
\end{eqnarray}
where $\tilde{\delta}(\ell)$ is the 2-D Fourier transform of the
radially symmetric halo profile:
\begin{equation}
  \tilde{\vec{\delta}}(\ell)=
  2\pi\int_0^\infty\d\theta\,\theta\,J_0(\ell\theta)\,\vec{\delta}(\theta)\;.
\end{equation}
The window $F(x)$ defines the smoothing window (also radially
symmetric) of our lens-plane pixels, which for circular pixels is
defined in Eq. \Ref{eq:window}.

For the scope of this analysis, the original halo signal,
$\vec{\tilde{\delta}}(\ell)$, is a vector of mode amplitudes on the
lens-planes, being zero except for the plane on which the halo is
located.  After filtering, l.h.s. of Eq. \Ref{eq:pixelsignal0}, the
signal will be spread along the line-of-sight, i.e. spread over the
output vector $\vec{\bar{\delta}}(\Theta_{\rm s},\alpha)$, as already
seen for the Fourier modes, Fig. \ref{fig:response}. Moreover, as
everything is linear, the response of a sum of haloes is just the sum
of the individual responses.

There is also a transverse (spread inside same lens-plane) point
spread function (PSF) associated with the filter, which can be
evaluated at separation $\Delta\theta$ from the halo centre by
offsetting the position of $\delta(\vec{\theta})$ in Fourier space,
\begin{equation}
  \vec{\tilde{\delta}}(\ell)\mapsto
  \vec{\tilde{\delta}}(\ell)\,
      {\rm e}^{{\rm i}/2(\vec{\theta}\vec{\ell}^\ast+\vec{\theta}^\ast\vec{\ell})}\;,
\end{equation}
and reassessing Eq. \Ref{eq:pixelsignal0} (for a radially symmetric
profile):
\begin{equation}\label{eq:filterpsf}
  \vec{\bar{\delta}}(\Theta_{\rm s},\alpha,\Delta\theta)=
  \frac{1}{2\pi}\int_0^\infty\d\ell\,\ell\,F(\ell
  \Theta_{\rm s})\,J_0(\ell\Delta\theta)\,\widetilde{\mat{W}}
  \vec{\tilde{\delta}}(\ell)\;.
\end{equation}
Obviously, the transverse PSF is symmetric (only a function of the
modulus of $\vec{\Delta\theta}$) and will, therefore, not bias the
angular position of a halo. Moreover, usually a strongly down-tuned
Wiener filter will be used, $\alpha\ll1$, which will make the
transverse spread relatively small. As a matter of fact, we will have
to apply an additional transverse smoothing to get a reasonable
signal-to-noise data cube. Conversely, the radial PSF will be of more
concern.

\begin{figure}
  \begin{center}
    \psfig{file=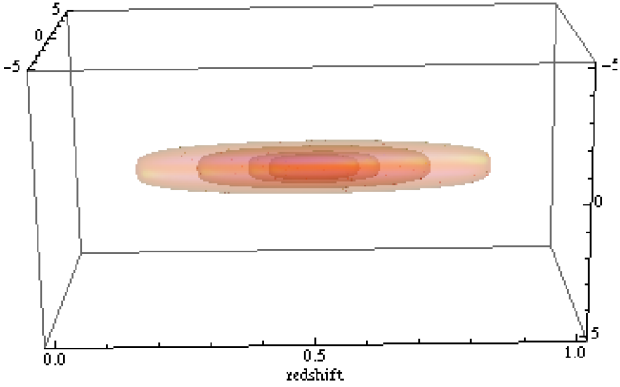,width=56mm,angle=-90}
  \end{center}
  \caption{\label{fig:toyresponse} Toy model example of the response
    to a halo with some density profile sitting at a redshift of
    $z=0.5$, originally with no radial extension. Plotted are four
    different iso-density surfaces with decreasing iso-densities,
    growing bigger in size for smaller iso-density values. Clearly,
    the surfaces become stretched -- the more, the smaller the
    iso-density -- in radial direction in a cigar-like manner (we
    loose redshift information) due to the response of the Wiener
    filter. As shown in Fig. \ref{fig:radialpsf}, the effect becomes
    smaller for smaller $\alpha$ (less Wiener prior) which, however,
    will increase the noise in the reconstruction. The iso-surfaces
    responding to structures situated at the redshift boundaries of
    the survey will be radially stretched in a asymmetric manner (not
    shown).}
\end{figure}

Fig. \ref{fig:toyresponse} gives a toy model example of the response
of the Wiener filter to some halo with profile $\delta(\theta,z_{\rm
h})$, sitting at a singular redshift $z_{\rm h}=0.5$. If the response
is a direction-independent function $C(z_{\rm h},z)$ -- in a toy model
fashion a Gaussian with mean $z_{\rm h}$ and width $\sigma_{\rm
z}=0.2$ -- then we will find in the reconstruction the 3-D map
$\delta(\theta,z_{\rm h})C(z_{\rm h},z)$. The general effect is that
the iso-density surfaces of the profile are stretched
radially. Although all $\delta(\theta,z_{\rm h})$ are being stretched
with the same $C(z_{\rm h},z)$, \emph{surfaces} with smaller
iso-densities are stretched more than higher iso-density
values. Qualitatively, this toy model describes quite well what we
find in the real reconstructions. Note that the surfaces are centred
about the true halo redshift since no $z$-shift due to the response
$C(z_{\rm h},z)$ was assumed for this illustration.

\begin{figure*}
  \begin{center}
    \psfig{file=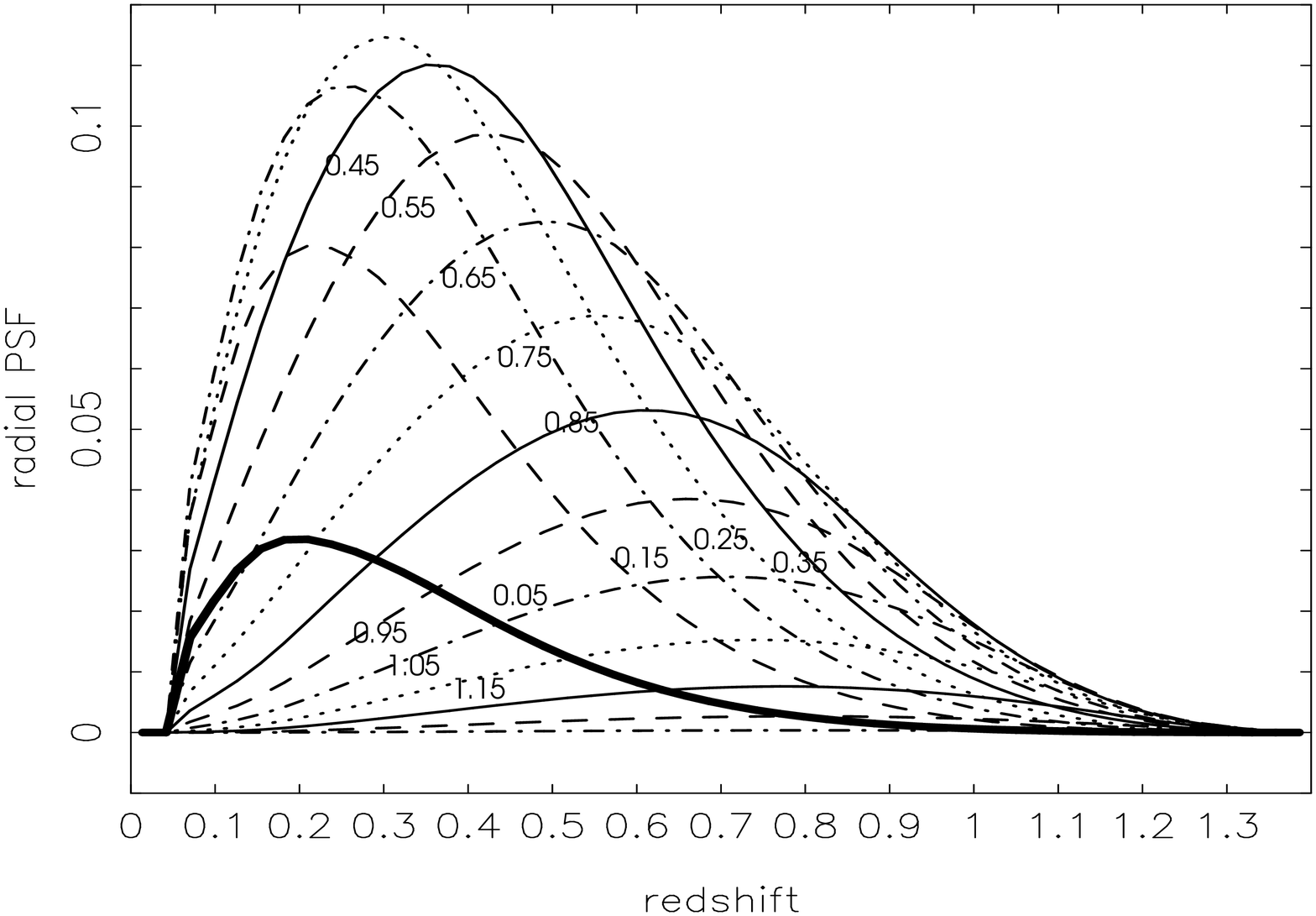,width=80mm,angle=0}
    \psfig{file=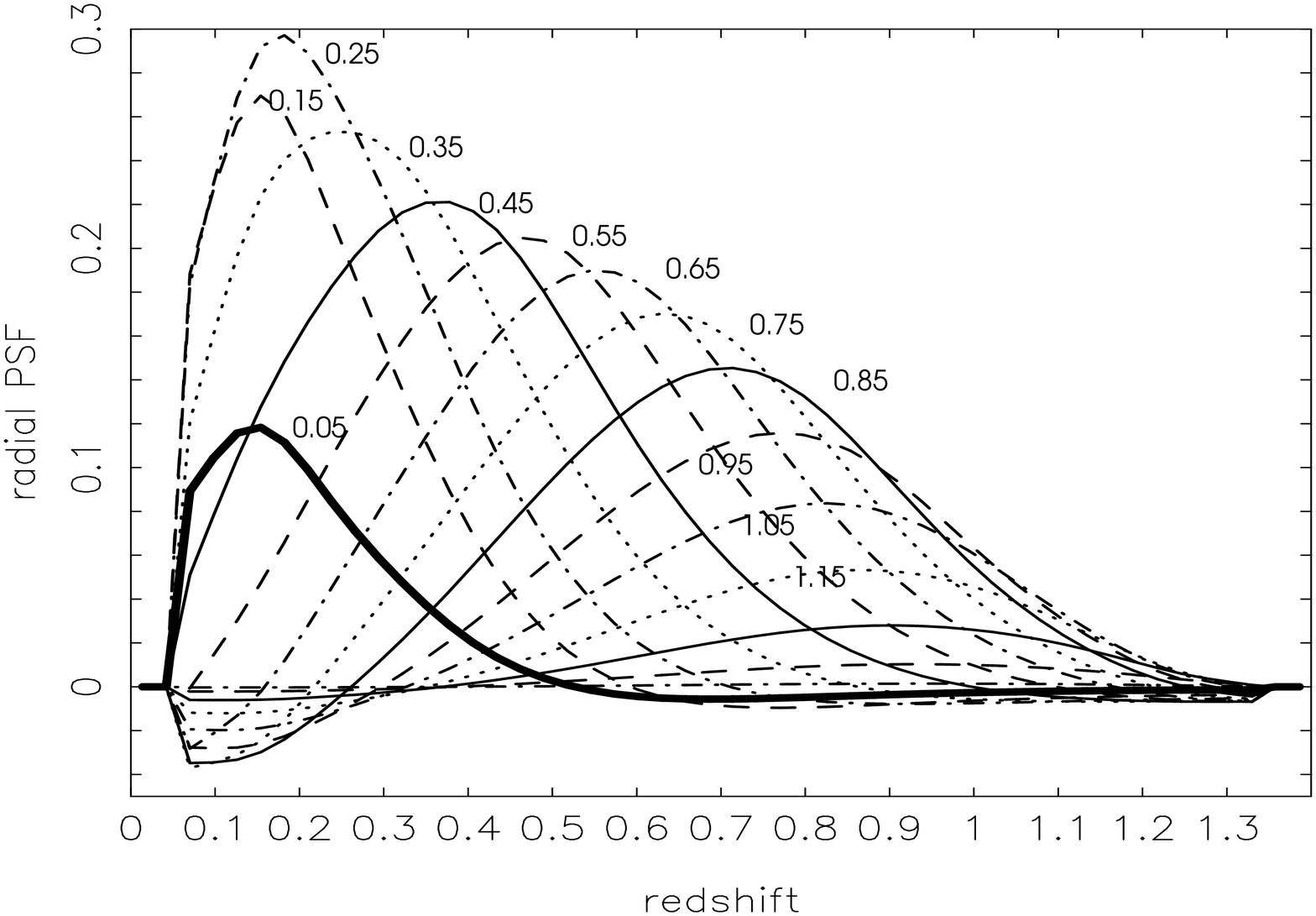,width=80mm,angle=0}
  \end{center}
  \caption{\label{fig:psfexample} The radial PSF of a fiducial survey
    as in Fig. \ref{fig:snpower} ($\bar{n}=30\,\rm arcmin^{-2}$; curves
    continue up to redshift $z=1.5$). A density pixel of radius
    $1^\prime$ sitting on a lens-plane with certain redshift (numbers
    attached to curves) is in the process of a reconstruction
    attenuated and spread out. The \emph{left figure} assumes a
    transverse Wiener filter with $\alpha=10^{-2}$, whereas the
    \emph{right figure} uses a radial Wiener filter with
    $\beta=10^{-3}$. For example, a pixel at $z=0.05$ is shifted to
    $z\sim0.2$ (maximum of PSF) and attenuated with a factor of
    $\sim0.03$ with transverse filtering (thick solid line). Note that
    the PSF of the radial filter can become negative.}
\end{figure*}

Under the idealised circumstances the radial point spread function
will be the same in all directions.  Since, in reality, we will have a
varying source number density as function of direction and a finite
survey area we can expect a more complex, direction dependent response
of the Wiener filter which accounts for inhomogeneous source
ellipticity noise patterns.

Assuming as a lowest-order approximation a homogeneous noise pattern
of the sources, we can work out the expected radial PSF of the Wiener
filter, i.e. the radial convolution $\mat{C}$, by means of
Eq. \Ref{eq:pixelsignalcirc}:
\begin{equation}\label{eq:radialpsf}
  \mat{C}= \frac{1}{2\pi}\int_0^\infty\d\ell\,\ell\,|F(\ell \Theta_{\rm s})|^2\,
  \widetilde{\mat{W}}\;.
\end{equation}
Here, an original pixel on the lens plane has the same shape and size
as a reconstructed pixel, i.e. $\tilde{\delta}(\ell)\propto F(\ell
\Theta_{\rm s})$. A pixel on the $i$th lens plane is spread over the
other lens planes according to $\mat{C}\vec{e}_i$, $\vec{e}_i$ is a
unity vector non-zero only at the original position of the pixel, or
put in another way: $C_{ij}$ is the signal-leakage of pixels on the
$i$th lens-plane into pixels on the $j$th lens-plane.  In an ideal,
unbiased reconstruction, $\widetilde{\mat{W}}=\mat{1}$, the matrix
kernel of the integral would be the unity matrix so that
\begin{equation}
   \mat{C}=
   \frac{1}{2\pi}\int_0^\infty\d\ell\,\ell\,|F(\ell \Theta_{\rm s})|^2\,\mat{1}=\mat{1}
\end{equation}
for a properly normalised pixel window function $F$. Thus, the
pixelised reconstruction would yield exactly the original matter
distribution on the lens-planes smoothed with the pixel window. In
reality, however, we will find a situation more like the one shown in
Fig. \ref{fig:psfexample}.

\begin{figure*}
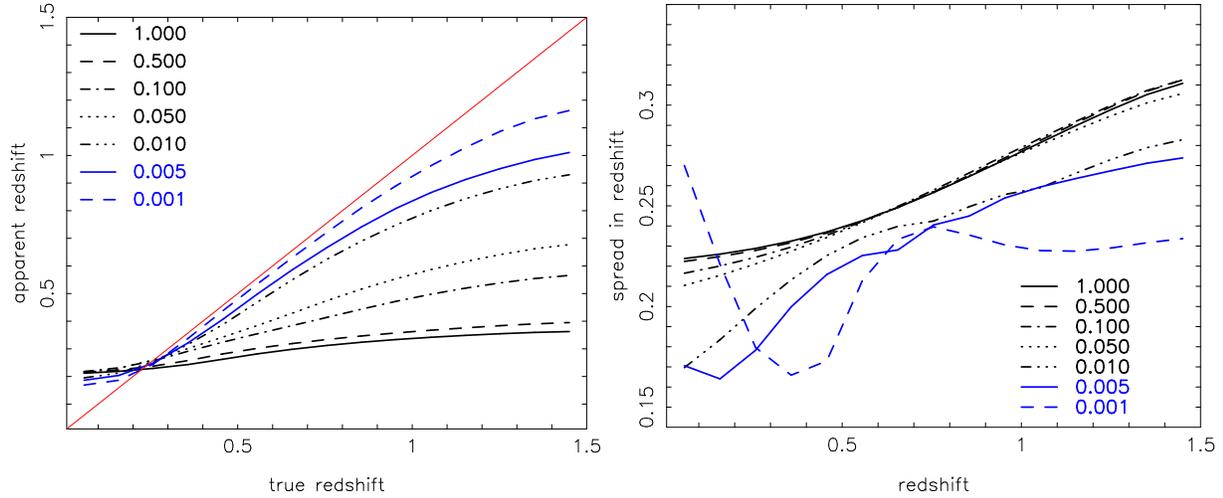

  \begin{center}
    \psfig{file=halo.zshift.ps, width= 65mm, angle=-90}
    \psfig{file=halo.zspread.ps,width=65mm, angle=-90}    
  \end{center}
  \caption{\label{fig:radialpsf} These plots quantify the shape of the
  radial PSF for circular pixels of radius $1^\prime$: the redshift of
  the maximum response (\emph{left panel}) and the z-width ($1\sigma$)
  of the PSF (\emph{right panel}) as function of redshift of the
  original pixel. The shape of the PSF also depends on the tuning
  parameter $\alpha$ (different line styles). See
  Fig. \ref{fig:psfexample} for a concrete example of the PSFs. The
  plots are based on a fiducial $\Lambda$CDM model with WMAP3-like
  parameters, an intrinsic source shape noise of
  $\sigma_\epsilon=0.3$, a source number density of $\bar{n}=30\,\rm
  arcmin^{-2}$ and a source redshift distribution as in
  Eq. \Ref{eq:fiducialpz}. The sources are binned between
  $z=0\ldots1.5$ in equally spaced bins with $\Delta z=0.05$.  To
  obtain the corresponding plots for different
  $\sigma^\prime_\epsilon$ or $\bar{n}^\prime$, but all other
  parameters kept, one has to rescale
  $\alpha\mapsto\alpha\times\left(\frac{\sigma^\prime_\epsilon}{\sigma_\epsilon}\right)^2\left(\frac{\bar{n}_{}}{\bar{n}^\prime_{}}\right)$,
  see Sect. \ref{sect:averagesn}.}
\end{figure*}

Now, Fig. \ref{fig:radialpsf} uses the results of the previous
discussion to quantify the quality of a 3-D reconstruction for a
particular fiducial cosmology and a realistic fiducial survey
employing a Wiener filtering scheme with $\alpha$-tuning. The plots
can be scaled to different $\sigma_\epsilon$ or $\bar{n}$ as explained
in the caption. We can see in the plots that, the shift and spread of
the signal in the reconstruction due to Wiener filtering decreases if
the filter is tuned down, which, on the other hand, will decrease the
signal-to-noise. We also find that the redshifts of objects near the
``edges'' of the survey (here: $z\sim0$ and $z\sim1.0$) are mostly
biased, high redshift objects appear to be more affected. The bias
tends to shift low-z objects to apparently higher redshifts, while
high-z objects are moved in direction of small redshifts.

\begin{figure*}
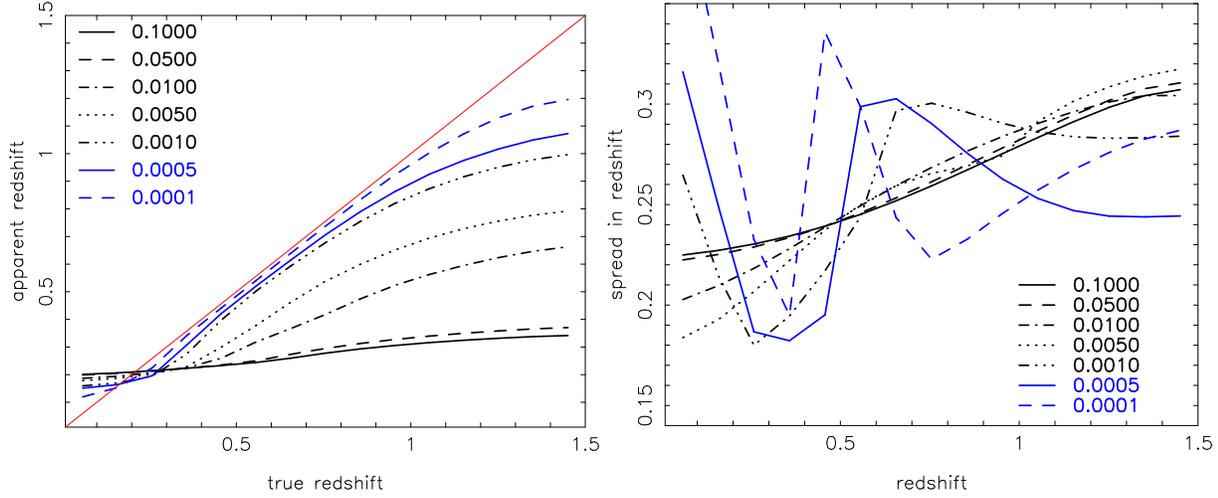

  \begin{center}
    \psfig{file=halo.zshift.radial.ps,width=65mm,angle=-90}
    \psfig{file=halo.zspread.radial.ps,width=65mm,angle=-90}    
  \end{center}
  \caption{\label{fig:radialpsfradial} Same as
    Fig. \ref{fig:radialpsf} but using a radial Wiener filter. For
    small $\beta$ the curves disperse in the right panel because the
    radial PSF starts to oscillate about zero similar to the filter
    without prior.}
\end{figure*}

In analogy to these plots, Fig. \ref{fig:radialpsfradial} shows the
same parameters but this time obtained with a radial Wiener filter. We
find a similar behaviour compared to the transverse filter, keeping in
mind that the tuning parameters of the radial and transverse filter
are not comparable: an $\beta$ of the radial filter ten times smaller
than the $\alpha$ of the transverse filter yields roughly the same
signal-to-noise (see next section). The expected signal based on the
radial filter gives, in contrast to the transverse our filter, a
different prediction as the ``true'' signal which is calculated from
Limber's equation. Looking more closely, one finds that a radial
filter might be used to slightly decrease the $z$-shift due to the PSF
($\beta\sim10^{-3}$ compared to $\alpha\sim10^{-2}$), although the
spread in radial direction may be larger. As this spread is partly
into the negative regime, see right Fig. \ref{fig:psfexample}, it may
not become immediately obvious but can have the effect of diminishing
structures along the line-of-sight -- or generating structures as a
response to an underdensity.

We noticed during the course of this work that a very simple radial
prior that gives all lens-planes the same prior,
$P_{ij}=\pi\Theta_{\rm s}^2$, gives often somewhat less biased
signal-to-noise maps so that it is worthwhile to compare in a concrete
case the benchmarks of this simple radial filter to the previous
radial filter.

As already pointed out in the introduction, we could in principle
deconvolve out reconstruction with the known PSF. This, unfortunately,
would completely remove the effect of the Wiener filter leaving us
with the extremely noisy and radially oscillating reconstruction of
the no-prior estimator.

\subsection{Signal-to-noise of matter haloes}
\label{sect:snhaloes}

Here we would like to compare the expected pixelised signal in a
reconstruction to the expected noise level within the same pixel.

If the density modes, $\tilde{\delta}(\ell)$, of Sect. \ref{sect:psf}
belong to a \emph{random field} -- such as the noise in the
reconstruction or the random density fluctuations in a particular
fiducial cosmological model -- the variance of the density contrast
inside the a pixel will turn out to be
\begin{equation}\label{eq:pixelwindow}
  \Ave{\bar{\delta}^2(\Theta_{\rm s})}^\prime=
  \frac{1}{2\pi}\int_0^\infty\d\ell\,\ell\,|F(\ell \Theta_{\rm s})|^2\,P(\ell)\;,
\end{equation}
for a power spectrum, $P(\ell)$, quantifying the fluctuations in
$\tilde{\delta}(\ell)$. This follows from Eq. \Ref{eq:smooth}.

If we apply this integral transformation to the power spectra
(matrices) Eqs. \Ref{fouriercovdelta} and \Ref{fourierpowdelta}, we
can transform the Fourier space representation to signal-to-noise
estimates inside pixels as function of lens-plane redshift. From that
one can infer that the average signal-to-noise inside a circular pixel
will be approximately the signal-to-noise of one effective
$\ell$-mode, Fig. \ref{fig:snpower}. As the integration kernel,
$x|F(x)|^2$, in Eq. \Ref{eq:pixelwindow} peaks at about $\ell
\Theta_{\rm s}\approx1.36$, the effective mode frequency will be near
$\ell_{\rm eff}\approx4666\,(\Theta_{\rm s}/1^\prime)^{-1}$.

The expected covariance of noise inside lens-plane pixels arranged
along the line-of-sight will be:
\begin{equation}\label{eq:pixelnoise}
  {\rm Cov}(\vec{\bar{\delta}}(\Theta_{\rm s},\alpha))=
  \frac{1}{2\pi}\int_0^\infty\d\ell\,\ell\,|F(\ell \Theta_{\rm
  s})|^2\,\tilde{\mat{N}}_{\rm MV}(\alpha)\;.
\end{equation}
This can be directly compared to the blurred signal from a single halo
(Eq. \ref{eq:pixelsignalcirc}), sitting at some redshift, being filtered
by our Wiener filter:
\begin{equation}
  {\rm\frac{S}{N}}\left(\bar{\delta}_i(\Theta_{\rm s},\alpha)\right)=
  \frac{\bar{\delta}_i(\Theta_{\rm s},\alpha)}{\sqrt{{\rm Cov}\left(\vec{\bar{\delta}}(\Theta_{\rm s},\alpha)\right)_{ii}}}\;.
\end{equation}
Note that we only consider the smoothed signal about the halo centre
here, where we expect to see the highest signal-to-noise of the halo
reconstruction. The Wiener filtering does not bias the angular
position of a halo (the phases of the Fourier modes), in particular
the centre position. Therefore, we will, on average, find the halo
centre in the reconstruction in the original direction as before
filtering.

For a singular isothermal sphere \citep[SIS;
e.g.][]{2001A&A...378..361B} we have
\begin{equation}
  \delta(\theta)= \frac{4\pi c^2a(w_{\rm
  h})}{3H_0^2\Omega_{\rm m}\Delta w}\left(\frac{\sigma_{\rm
  v}}{c}\right)^2\frac{1}{f_{\rm K}(w_{\rm h})\theta}\;,
\end{equation}
where $\sigma_{\rm v}$ is the velocity dispersion inside the SIS and
$a(w_{\rm h})$ the scale factor at the radial comoving distance of the
halo centre, $w_{\rm h}$. The Fourier transform of the SIS profile is
\begin{equation}
  \tilde{\delta}(\ell)=
  \frac{8\pi^2 c^2a(w_{\rm
      h})}{3H_0^2\Omega_{\rm m}\Delta w}\left(\frac{\sigma_{\rm
      v}}{c}\right)^2\frac{1}{f_{\rm K}(w_{\rm h})\ell}\;.
\end{equation}

As the signal inside a pixel, Eq. \Ref{eq:pixelsignalcirc}, is linear
in the original underlying amplitude of $\delta$, whereas the noise
inside the pixel is independent of $\delta$ (Eq. \Ref{eq:pixelnoise})
we can already infer that the signal-to-noise of the SIS in the
reconstruction has to scale with ${\rm S/N}\propto\sigma^2_{\rm v}$,
roughly the mass of the halo. We can expect this to be approximately
true for more realistic halo profiles as well.

More specifically, \citet{2001A&A...378..361B} give for the virial
mass\footnote{The mass contained in a radius at which the mean density
of the SIS equals $200$ times the critical density of the Universe.}
of the SIS
\begin{eqnarray}
  M_{200} &=& \frac{2^{3/2}\sigma_{\rm v}^3}{10GH(z)}\\
  &\approx&
  6.58\times10^{14}\msol\,h^{-1}\,
  \left(\frac{\sigma_{\rm v}}{10^3\,{\rm kms^{-1}}}\right)^3
  \frac{1}{E(z)}\;,
\end{eqnarray}
where 
\begin{equation}
  E(z)\equiv
  \sqrt{\Omega_{\rm m}(1+z)^3+(1-\Omega_{\rm m}-\Omega_\Lambda)(1+z)^2+\Omega_\Lambda}
\end{equation}
is the normalised Hubble parameter, $H(z)/H_0$, as function of
redshift and $G$ Newton's constant. For $\Omega_{\rm m}=0.23$ and
$z<1$, $E(z)\approx1-0.345\,z$ provides a fair
approximation. Therefore, the scaling for a SIS is ${\rm S/N}\propto
M_{200}^{2/3}$.

\begin{figure*}
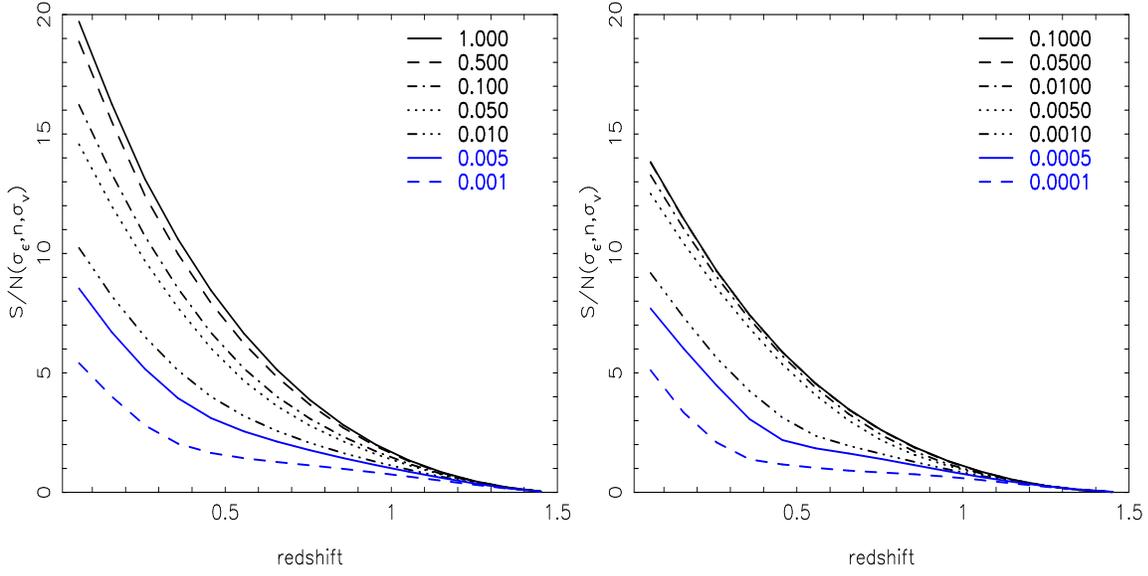

  \begin{center}
    \psfig{file=halo.signal-to-noise.ps,width=75mm,height=75mm,angle=-90}
    \psfig{file=halo.signal-to-noise.radial.ps,width=75mm,height=75mm,angle=-90}
  \end{center}
  \caption{\label{fig:sissignaltonoise} Signal-to-noise of a
    reconstructed SIS, positioned at different redshifts, $x$-axis,
    with $\sigma_{\rm v}=10^3\,\rm km s^{-1}$; this corresponds to
    $M_{200}=6.6\times10^{14}\,\msol h^{-1}$
    ($1.6\times10^{15}\,\msol h^{-1}$) at $z=0$ ($1$). Plotted is as
    function of tuning parameter (line styles) the signal-to-noise at
    maximum response. The assumed fiducial survey has the parameters
    outlined in Fig. \ref{fig:radialpsf}. To obtain the corresponding
    plots for different $\sigma^\prime_\epsilon$, $\sigma^\prime_{\rm
      v}$ or $\bar{n}^\prime$, but all other parameters kept, one has
    to rescale
    $\alpha\mapsto\alpha\times\left(\frac{\sigma^\prime_\epsilon}{\sigma_\epsilon}\right)^2\left(\frac{\bar{n}_{}}{\bar{n}^\prime_{}}\right)$
    and ${\rm \frac{S}{N}}\mapsto{\rm
      \frac{S}{N}}\times\left(\frac{\sigma_{\rm v}}{\sigma_{\rm
	v}^\prime}\right)^2\left(\frac{\sigma_\epsilon}{\sigma_\epsilon^\prime}\right)\left(\frac{\bar{n}_{}}{\bar{n}^\prime_{}}\right)^{1/2}$,
    as described in Sect. \ref{sect:averagesn}. \emph{Left figure:}
    transverse Wiener filter. \emph{Right figure:} radial Wiener
    filter.}
\end{figure*}

Fig. \ref{fig:sissignaltonoise} displays the signal-to-noise of the
central pixel of a smoothed, reconstructed SIS. The calculations are
done for a SIS with fixed velocity dispersion but can easily be
rescaled for other $\sigma_{\rm v}$, or different fiducial surveys
with varied source density or shape noise. See the caption of the
figure for details.

The signal-to-noise of the central pixel is highest for small halo
redshifts and decreases rapidly towards higher redshifts. This is
because the number of sources behind the halo decreases as we move
towards higher halo redshifts. Therefore, the technique identifies
haloes most efficiently at low redshifts and becomes increasingly
ineffective towards higher redshifts. We also found by varying the
redshift distribution of the sources, while keeping their number
density and intrinsic ellipticity distribution constant, that the
signal-to-noise fall-off is shallower for a deeper survey, although
the $z$-spread increases owing to the same number of sources being
distributed over a larger $z$-range.

We also gather from the results, that the signal-to-noise decreases as
we tune to smaller values of $\alpha$, thereby reducing the Wiener
regularisation.  For a transverse filter and the given fiducial
survey, values of $\alpha\lesssim10^{-2}$ are necessary to reduce the
bias of the Wiener filter ($1\sigma$-$z$-spread and $z$-shift less
than $\sim0.2$) to an acceptable level, Fig. \ref{fig:radialpsf}. For
our radial filter, a value of $\beta\lesssim10^{-3}$ should be
used. We excluded values smaller than $\alpha\leq10^{-3}$ (radial:
$\beta\leq10^{-4}$) as this brings us too close to the no-prior
filter which exhibits heavy, undesired oscillations in radial
direction and very poor signal-to-noise. Note that $\alpha$ also
depends on the redshift distribution of the sources.

\begin{figure*}
   \begin{center}
     \psfig{file=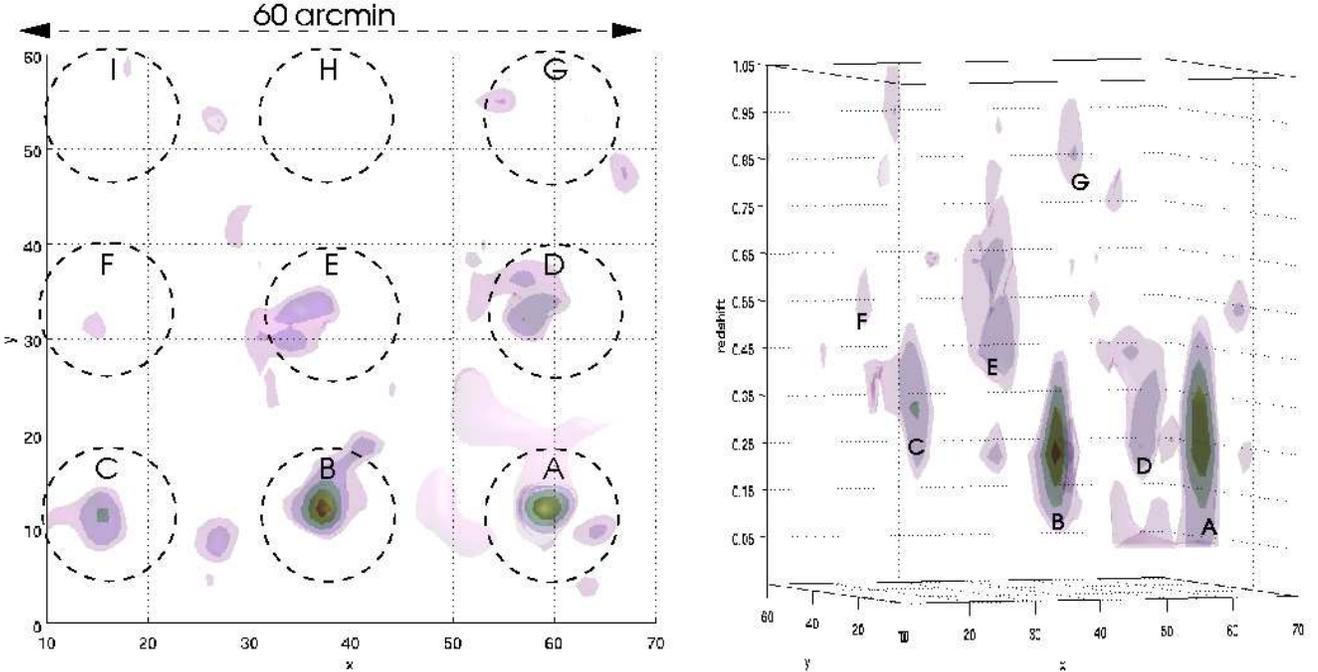,width=175mm,angle=0}
   \end{center}
   \caption{\label{fig:sismock} A set of nine SISs ($\sigma_{\rm
   v}=10^3\,\rm km\,s^{-1}$) with redshifts $z=0.1,0.2,\ldots,0.9$
   (A-I in alphabetical order) has been planted into a particular
   noise realisation of the fiducial survey defined in
   Fig. \ref{fig:radialpsf}. The left panel is a 2-D projection of the
   reconstruction onto the sky while the right panel is a projection
   showing the extension of the volume in redshift. The differently
   coloured contours are constant signal-to-noise levels corresponding
   to ${\rm S/N}=8,6,5,3,2.5$ (darkest to brightest colour). The
   reconstruction is smoothed with a kernel of radius $1^\prime$. The
   employed transverse Wiener filter has $\alpha=0.01$. Statistical
   errors can dissolve peaks as can be seen for D and E or shift peaks
   towards wrong redshifts, see D for instance. Note that all
   redshifts are biased to some extend, left panel of
   Fig. \ref{fig:radialpsf}.}
\end{figure*}

An example reconstruction (transverse filter) of a set of SISs planted
into a one-square-degree field-of-view between redshift $z=0...1$ can
be found in Fig. \ref{fig:sismock}. This example underlines the
theoretical expectation for our fiducial survey that the
signal-to-noise of the haloes in the reconstruction drops quickly if
we move towards higher redshifts; the most distant haloes (F-I) are
barely or not distinguishable from the background noise. Note that the
signal-to-noise levels cited in the figure caption were computed based
on $500$ FFT-reconstructed noise realisations, see Appendix
\ref{sect:covariance}. One can also observe the effect of the PSF that
spreads out the peaked matter distributions and moves, for instance,
the maximum of the response of low-redshift haloes towards higher
redshifts. Due to the noise in the data, some peaks (C and D) fragment
into different parts and turn up at clearly wrong redshifts. An
analogous reconstruction ($\beta\sim10^{-3}$) employing a radial
filter looks essentially the same.

\subsection{Edge effects}

\begin{figure}
   \begin{center}
     \psfig{file=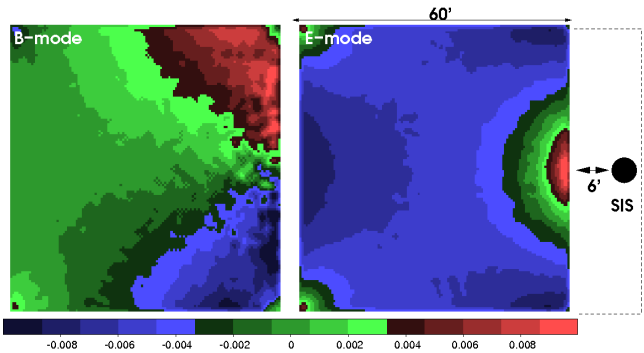,width=85mm,angle=0} 
   \end{center}
   \caption{\label{fig:edgepic} Reconstruction of the lensing
     convergence on a $60^\prime\times60^\prime$ patch falsely assuming
     that all shear signal originates from matter inside the patch
     (\emph{left}: B-mode, \emph{right}: E-mode). Actually, the signal
	 is solely produced by a SIS, $\sigma_{\rm
	   v}=10^3\,\rm km\,s^{-1}$ and $z=0.2$, sitting in a separation of
	 $6^\prime$ off the right edge of the patch. The noise in the
	 images is due to the finite number density of (intrinsically
	 circular) sources, $\bar{n}=30\,\rm arcmin^{-2}$, used for the
	 fiducial survey here.}
\end{figure}

The reconstruction algorithm assumes that the cosmic shear signal
within the patch of the survey originates solely from matter within
the field-of-view. This is the same for all Kaiser-Squires type
algorithms \citep[KS;][]{1993ApJ...404..441K}. This introduces,
especially at the boundaries of the field, a bias if the actual source
of the shear within the field is outside the patch. This problem would
not exist for a full-sky implementation of the algorithm.

To have an qualitative assessment of the impact of this bias, we
consider one particular, extreme case which is shown in
Fig. \ref{fig:edgepic}. For this mock survey, a square patch with
$60^\prime$ on each side and the previous redshift distribution of
sources, we made up the case that there are no sources of shear at all
inside the patch area. The total shear signal observed stems from one
massive singular isothermal sphere located close to the right edge of
the patch. The figure shows a KS reconstruction (left: B-mode, right:
E-mode) of the convergence inside the patch ignoring an outside source
as possible origin of the shear. The 3-D reconstruction of the density
contrast for a lens plane at about the redshift of the SIS exhibits
exactly the same pattern. Lens planes well separated from the SIS
redshift show no or little reconstructed mass density while
neighboured lens planes may be affected due to the radial spread of
the reconstruction PSF.

It can be seen from this test that a source outside the observation
area can indeed produce mass density inside the reconstruction
patch. However, this ``ghost signal'' seems to be mainly focused on
the edges. The largest inferences can be seen on the edge closest to
the SIS, namely at the middle and at the corners, and on the edge
opposite to that with roughly the same but inverted signal
pattern.

Furthermore, it can be seen that this ghost signal is associated with
a B-mode signal of roughly the same amplitude than the E-mode signal,
also most prominent at the edges (closest and opposite) but leaking
somewhat more into the patch. Therefore, if an off-patch source
produces a significant signal inside the patch area it may reveal
itself by an equally significant B-mode signal.

Note that the response pattern looks different if we use FFTs for the
reconstruction, as they assume periodic boundary conditions. With FFT
the SIS can actually be reconstructed as mirror image near the edge
opposite to the edge closest to the true position of the SIS.  Yet, we
find that even with periodic boundary conditions a B-mode signal is
generated inside the patch which has the same order of magnitude than
the E-mode signal.

This bias can be controlled by making the reconstruction area larger
than the patch area, attributing to this additional patch frame
infinite noise. By doing this, we explicitly allow the reconstruction
algorithm to place sources of shear outside the observed patch area.

\subsection{Reduced shear}

 Throughout this paper we assumed the weak lensing approximation to be
perfectly valid, i.e. ellipticities of galaxy images are unbiased
estimators of the cosmic shear and not the reduced shear,
$\gamma/(1-\kappa)$, as they actually are.

If the distinction between reduced shear and shear is ignored, the
convergence will in the vicinity of density peaks biased towards
smaller values and hence the density will be biased towards smaller
values as well.

Presumably, the algorithm can be modified in this respect by running
the linear algorithm on $\epsilon^{(i)}_j(1-\kappa^{(i)}_j)$, where is
$\vec{\kappa}=\mat{Q}\vec{\delta}$ is the minimum variance convergence
reconstruction of the previous run; for the first run one starts off
with $\kappa^{(i)}_j=0$. For a 2-D convergence reconstruction this
iterative approach is known to converge quickly to the right solution
\citep[][Sect. 4.3]{1998ApJ...506...64S}. The noise of the nonlinear
reconstruction would have to be estimated by a series of noise
realisations. We will postpone a test of this approach for a
3-D reconstruction to a future paper.

\section{Outlook}
\label{sect:mockdata}

\begin{figure*}
   \begin{center}
     \psfig{file=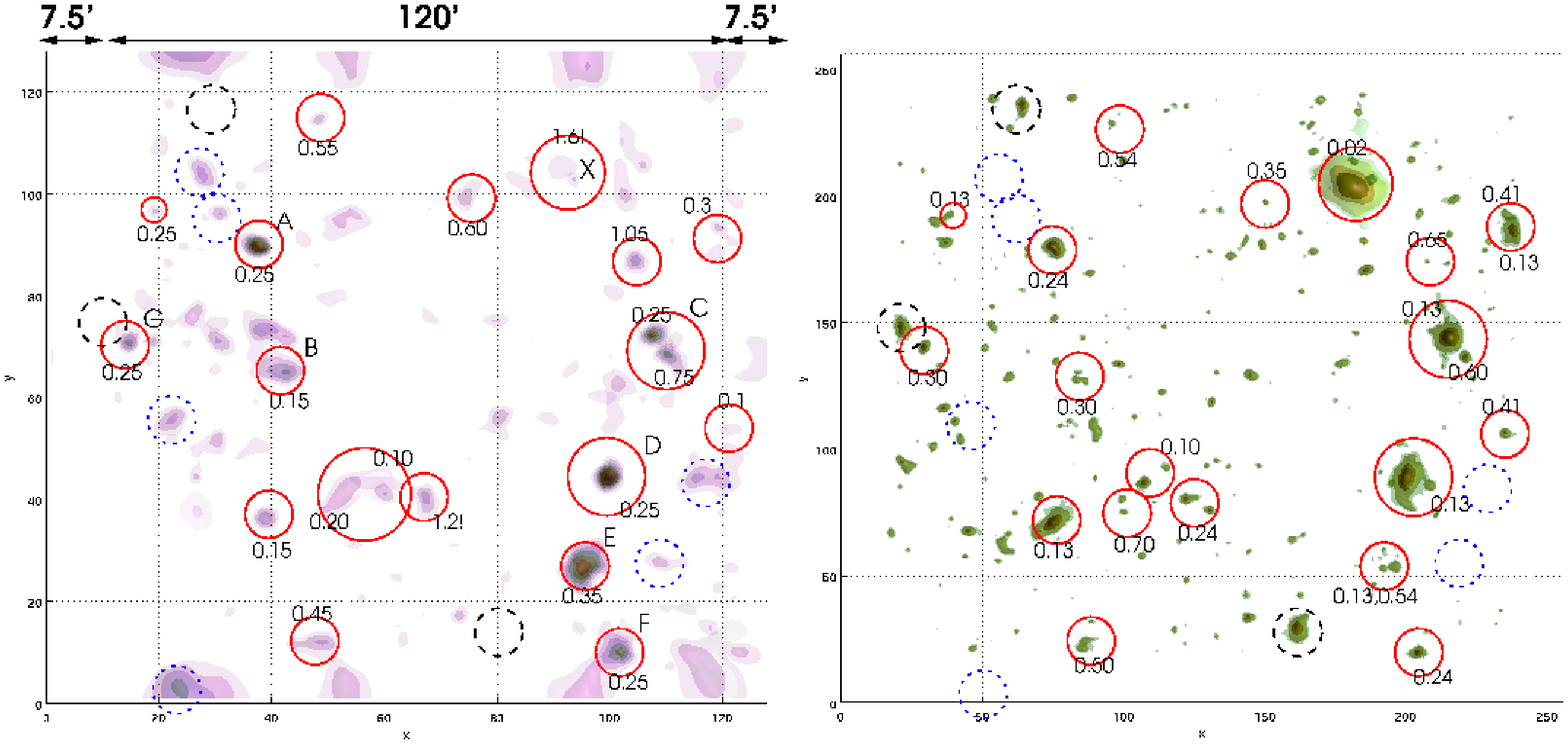,width=175mm,angle=0}
     \psfig{file=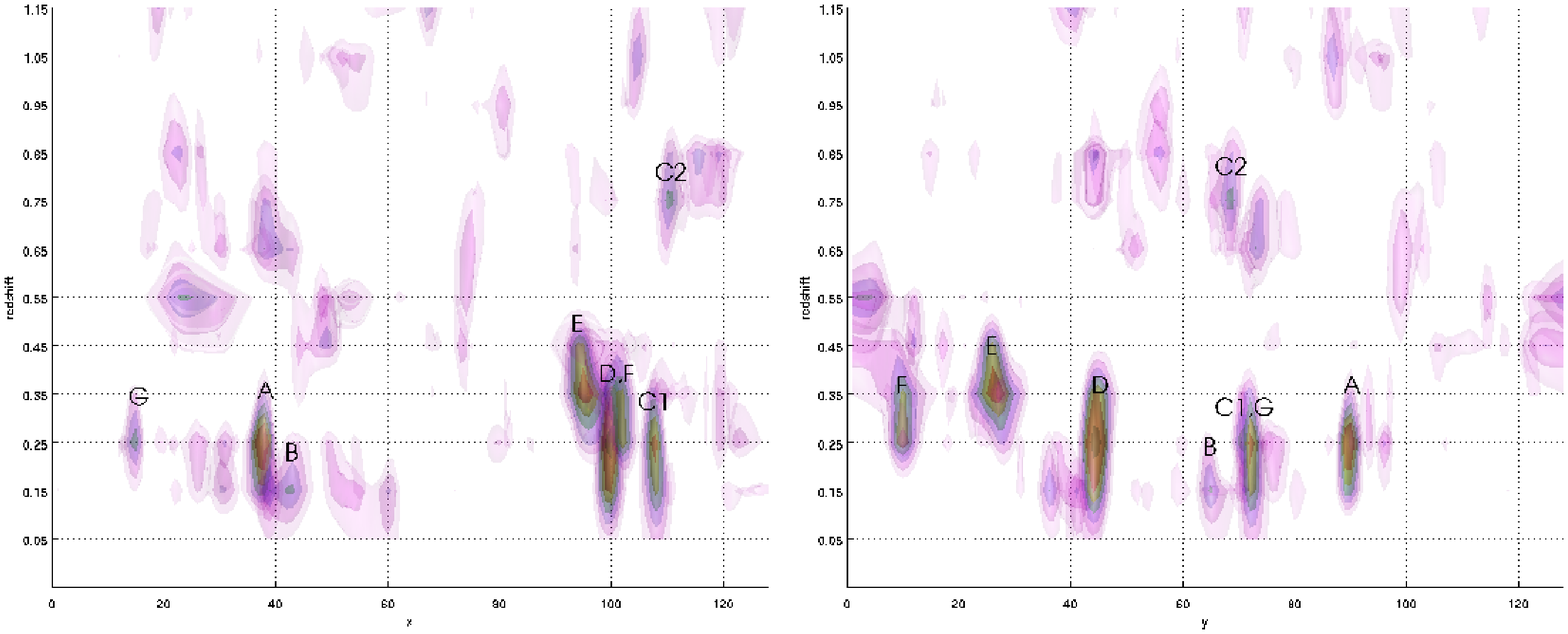,width=175mm, angle=0}
   \end{center}
   \caption{\label{fig:nbodypic} 3-D mass density reconstruction of a
     $2\,{\rm deg}\times2\,{\rm deg}$ patch of a mock lensing survey
     with $100\,{\rm arcmin}^{-2}$ source number density and mean
     source redshift $\bar{z}=1.1$ with \mbox{$z\le2.0$}. The patch
     was chosen to have a large number of high-density peaks. The
     intrinsic shape variance is chosen to be
     \mbox{$\sigma_\epsilon=0.3$}. The reconstruction was done on a
     $128\times128\,\rm pixel^2$ grid with $25$ source $z$-bins
     ($\Delta z=0.08$), and $20$ lens planes ($\Delta z=0.1$). The
     reconstruction was done on grid larger than the observed area,
     allowing for a frame with $7^\prime\!\!.5$ width. A transverse
     Wiener filter with $\alpha=0.05$ was employed. For the
     signal-to-noise maps, which are shown here, the data was
     randomised $100$ times. The upper left (projection on the sky;
     $x$ vs $y$), lower left ($x$ vs. redshift) and lower right panel
     ($y$ vs. redshift) show the signal-to-noise data cube of the
     reconstruction with 3-D iso-levels of \mbox{${\rm
     S/N}=6,5,4.5,4,3.5,3,2.5$}. The upper right panel shows for
     comparison true high over-densities (not over-densities due to
     projection) on the lens-planes used for the ray-tracing. The most
     significant peaks, ${\rm S/N}\ge4$, can be identified and matched
     with true matter peaks (solid circles). Several less significant
     peaks can be matched as well (not all encircled). But there are
     also peaks in the reconstruction with no apparent real
     counterpart (dotted circles) or high density peaks in the true
     distribution with no counterpart in the reconstruction (dashed
     circles); not all cases have been highlighted. The numbers denote
     the estimated (left) or true redshifts (right) of the
     structures. The most significant peaks in the upper left panel
     are labelled with alphabetic letters, A-G, and can be identified
     in the radial projections in the lower row. Label $X$ denotes a
     case discussed in the text. \changed{No attempt has been made to
     correct for the $z$-shift bias which is most evident for
     $z\lesssim0.15$.}}
\end{figure*}

For a demonstration and an outlook of what may be achieved with this
technique in the near future we apply the reconstruction method to a
simulated lensing survey patch of four square degree size. The
simulated survey has a source redshift distribution as in
Eq. \ref{eq:fiducialpz} but now with $z_0=1.0$, a source number
density of $\bar{n}=100\,\rm arcmin^{-2}$, corresponding to
$1.6\times10^6$ sources, and an average intrinsic ellipticity variance
of $\sigma_\epsilon=0.3$ for all sources. It is further assumed that
we have (photometric) redshifts for all individual galaxies and that
we are unable to obtain (photometric) redshifts beyond a redshift of
$z=2$, whence we truncate $p(z)$ at a redshift of two. The mean
redshift of the sources is therefore $\bar{z}=1.1$. These survey
parameters roughly reflect a space-based lensing survey which we
consider as the best possible choice for mass reconstructions due to
the small ratio $\sigma_\epsilon/\sqrt{\bar{n}}$, which is the main
factor in the signal-to-noise of the maps.

The mock survey was generated by ray-tracing through an $N$-body
simulation run with the publicly available version of GADGET-2
\citep{GADGET-2}. We have used $256^3$ Dark Matter particles in a box
with a side length of $150\,h^{-1}\,{\rm Mpc}$, which together with
the adopted $\Lambda$CDM cosmology ($\Omega_{\rm m}=0.25,\Omega_{\rm
b}=0.04, \Omega_\Lambda=0.75, \sigma_8=0.78$) results in a particle
mass of $1.2\times 10^{10}\,h^{-1}\,M_\odot$.  We have traced $2048^2$
light rays trough 25 matter slices with a thickness of
$150\,h^{-1}\,{\rm Mpc}$ up to $z=2$ using the standard
multiple-lens-plane algorithm \citep[e.g.][and references
therein]{mrrt}. This yields the Jacobian matrix of the lens mapping to
the back side of each matter slice. We have then created the mock
survey by randomly sampling the slice boundaries with galaxies
according to the desired redshift distribution, onto which we
interpolate the shear from the light ray positions.

For the reconstruction, we subdivide, equally in redshift, the source
catalogue into $25$ sub-samples with width $\Delta z=0.08$. This
number of source bins is fixed here by the number of lens planes
originally used for ray-tracing to obtain the simulated data. In
reality, $\Delta z$ reflects the average uncertainty in the redshift
estimates which is probably somewhat too pessimistic with
$0.08$. However, we would like to stress that a much finer binning
does not make much difference in the reconstruction -- at least if the
redshift estimates are not biased and the redshift distribution inside
the bins is accurately known -- since the lensing efficiency is but a
slowly changing function with source redshift. This is related to the
known fact that increasing the number of source bins beyond a few, in
constraining cosmological parameters with lensing tomography, does not
significantly increase the constraints. Hence, we do not expect a
notable improvement by using a larger number of source bins.

The number of lens planes, on which the matter density contrast is to
be estimated, is set to twenty ranging between $z=0\ldots2$ with
$\Delta z=0.1$. We do not use more lens planes, as a radial resolution
exceeding $\Delta z=0.1$ should not be expected according to the
foregoing discussion. In the following plots, we interpolated between
the lens planes.

Fig. \ref{fig:nbodypic} shows the reconstruction of the simulated
patch as signal-to-noise map either projected onto the sky or in two
different radial projections. Note that the sky-projections are
actually 3-D signal-to-noise maps. A signal-to-noise map of the
projected matter distribution, as commonly extracted from lensing
surveys (smoothed convergence maps), can be anticipated to have higher
significance.  For comparison also a sky-projection of the most
massive density peaks on the original lens planes is displayed. For
this demonstration a patch particularly rich in over-density peaks was
chosen, on average the number of peaks in a four square degree patch
in a WMAP3-like universe should be expected to be smaller.

We observe that although we have a quite optimistic lensing survey,
the significant features in the reconstruction mainly correspond to
the most massive haloes in the Dark Matter density field. A detection,
$\rm S/N\gtrsim3$, is restricted to the regime $z\lesssim1$. Below
$\rm S/N\approx3.5$ we find cases in the reconstruction which do not
seem to have a real counterpart. Those can be explained as pure
statistical flukes of the noise pattern which are still to be expected
on this signal-to-noise level. It may be that $100$ noise
realisations, as performed here, do not suffice to estimate the noise
pattern accurately enough everywhere. We also seem to have cases where
we have no significant detection of arguably high peaks in the true
Dark Matter density. The problem may be here that we are locally
lacking the required sampling of background sources. Peaks along the
same or close to the same line-of-sight are not always distinguished
from each other and merged into a mediate redshift instead (see E),
although there are cases where the algorithm succeeds in doing so (see
C).

Quite interesting is the case X, upper right quadrant, where we have a
large peak at very low redshift $z=0.02$ which however is completely
missed by the reconstruction or might be hinted at by a very weak
detection at a completely wrong redshift. The explanation for this
case probably is that the lensing efficiency for a structure at a
redshift that low is either too small for sources at higher $z$ or the
number density of sources which in principle would be sensitive enough
to light-deflections at that low redshifts is too small (low $z$). In
this sense, the lensing approach appears to be blind towards structure
at very low $z$.

Seemingly, the fiducial survey does not perform a lot better than the
inferior survey in Fig. \ref{fig:sismock}, where we assumed
$\bar{n}=30\,\rm arcmin^{-2}$ and a somewhat shallower mean
redshift. An increase in $\bar{n}$ from the latter to the former
survey, although observationally rather challenging, only improves the
average signal-to-noise by a mere factor of $\sqrt{10/3}\sim1.8$. This
again highlights the problem of all lensing mapping schemes.

The signal-to-noise of the reconstruction can be strongly increased by
using a more conservative, larger, tuning parameter (here:
$\alpha=0.05$). For instance, structure D can be boosted to a
signal-to-noise of approximately ten by using $\alpha=1$, however on
the expense of loosing information on the redshifts, revealed by a
significantly larger spread in radial direction, and more $z$-shift
bias.

Note that the $z$-shift bias is somewhat different in a
signal-to-noise data cube as opposed to a density contrast data cube
which was discussed in the foregoing section. Moreover, we have not
tried to correct this bias in our redshift estimates. \changed{In the
plot, therefore, particularly structures at low redshift,
\mbox{$z\lesssim0.15$}, are artificially shifted to larger redshifts,
$z\sim0.2$.}

The redshift estimates of structures more significant than $\rm
S/N\approx4$ are all in all reliable within $\pm0.1$, provided they
are not unresolved structures at different redshifts, but become
increasingly inaccurate below that signal-to-noise level. 

\high{The scenario discussed here contemplates a case where we have a
small but deep patch of which we would like to obtain a 3-D map with a
preferably good spatial resolution. Another scenario would be a
(almost) full-sky survey to which heavy smoothing is applied to study
the very large-scale density modes. Fig. \ref{fig:snpower} implies that
for those modes the reconstruction may become easier than for the
modes on a small patch as the signal-to-noise improves for lower
$\ell$ and the $z$-shift bias becomes less severe.}

\changed{The situation may be further improved if the shear data is
combined with higher-order lensing information, such as gravitational
flexion, that may be available in the future
\citep{2006MNRAS.365..414B,2007ApJ...660..995O}. In the weak lensing
regime, higher-order moments of lensing distortions can be modelled as
linear functions of the lensing convergence, as derivatives of
$\kappa$. Therefore, higher-order moments are easily implemented into
the algorithm as new extra set of redshift bins with a new linear
projection operator ``$\mat{P}_{{\rm F}\kappa}$'' as opposed to
$\mat{P}_{\gamma\kappa}$, $\mat{Q}$ remains unchanged. Very crudely --
same number of sources as for cosmic shear and same (uncorrelated)
signal-to-noise provided -- this will be equivalent to effectively
increasing the number density of sources in a conventional cosmic
shear survey by a factor of $N+1$, where $N$ is the number of new
lensing distortion moments that is included; the signal-to-noise in
the reconstruction consequently goes up by $\sqrt{N+1}$. Actually, as
already known from first 2-D applications
\citep{2006MNRAS.365..414B,2007ApJ...666...51L}, the 3-D maps will
become better on small angular scales with less improvement on larger
scales as the signal-to-noise of flexion-sampled mass density fields
is scale-dependent with most information on small scales. The radial
resolution of the maps, however, will only moderately benefit from the
new information, i.e. no more than from an extra cosmic shear
catalogue with same size, as the dependence of higher-order lensing on
redshift is identical to that of cosmic shear.}

\section{Summary and conclusions}

In this paper we presented a linear algorithm -- and two practical
ways of implementing it (Appendix
\ref{sect:kdtree},\ref{sect:fourierrecon}) -- that allows to
reconstruct the three dimensional distribution of matter based on a
lensing survey with redshift information. All available redshift
information, accurate source redshifts binned into thin slices and
broad redshift information such as wide redshift distributions for
faint, high-$z$ source subsamples, can be combined to find a matter
distribution that should be on average closest to the true
distribution. The statistical uncertainties of the source redshifts,
attached as PDF's of the redshifts to the corresponding source
subsample, and the statistical uncertainty of the shape measurement
can be properly factored into the reconstruction. Principally, as
shown in the paper, it is also possible to account for intrinsic
alignments of the sources, that have an impact on the expected shape
noise covariance entering the filter, and shear-intrinsic alignments.

The presented algorithm, being a generalisation of the algorithm
proposed by \citet{2002PhRvD..66f3506H}, yields the most probable
matter distribution, if the matter density field obeys Gaussian
statistics, or a minimum-variance solution otherwise. Due to the heavy
smoothing required in a 3-D matter distribution reconstruction based on
lensing tomography, the reconstruction is probably close to the
maximum-likelihood solution.

The algorithm is a Wiener filtering of the lensing tomography data.
The strength of the Wiener filter prior, i.e. the assumed second-order
correlations between densities in the reconstruction, can be tuned
with an additional parameter, $\alpha$. We looked into two different
cases for a regularisation, namely one case for which only
correlations between densities at the same redshift are given
(transverse filtering) and a case where only correlations along the
line-of-sight are specified (radial filter). Both priors show similar
properties, it is not possible to single out one filter as superior
filter.

We showed that a reconstruction with virtually no prior,
$\alpha\sim0$, gives an unbiased albeit extremely noisy
reconstruction. Moreover, a reconstruction with small $\alpha$
exhibits strong radial oscillations. The oscillations vanish and the
signal-to-noise improves considerably when one increases $\alpha$,
thereby enhancing the effect of the Wiener prior. However, a Wiener
filtered reconstruction represents, as all Wiener filter estimators, a
biased image of the true underlying signal. This bias can in our
context be expressed as a radial PSF or response of the reconstruction
for which we give an analytic expression in
Eq. \Ref{eq:radialpsf}. The effect of the PSF is that a peak in the
original density field will, on average, be stretched out and shifted
in radial direction, see Fig. \ref{fig:psfexample}. The stretch and
shift depend on the fiducial parameters of the survey and the tuning
of the filter. The bias essentially reflects our inability to pin down
exactly the redshift of a density peak due to the noise in the data.

Generally, the larger the $\alpha$-tuning, the better the
signal-to-noise of the reconstruction but the more biased becomes the
reconstruction. In the extreme case of over-regularisation, one
obtains essentially a 2-D reconstruction on the sky stretched out over
the entire radial range lacking any radial information, providing the
best signal-to-noise of a map of the projected matter distribution
though.

As strategy, we suggest to choose a parameter $\alpha$ with moderate
$z$-shift bias, see e.g. Figs. \ref{fig:radialpsf} and
\ref{fig:radialpsfradial}, and to relabel the redshifts of the
reconstruction lens-planes according to the expected statistical
average of the $z$-shift bias. At low redshift the radial PSF flattens
out, as can be seen in these figures (there: $z\lesssim0.3$). A
similar behaviour can be observed at the high redshift end of the
survey. In these regions, we are hence unable to uniquely relabel the
lens plane redshifts. At the low redshift range of that fiducial
survey, all we can say is that the structure generating a peak in the
reconstruction has to be located somewhere below $z\lesssim0.3$.

The reconstruction will be relatively noisy so that a final smoothing
in 2-D (lens planes only) or 3-D, for example with a Gaussian kernel,
has to be applied. The thereby obtained smoothed reconstruction is
still a minimum-variance reconstruction of the true density field
subject to the same smoothing, an invertible smoothing provided.

As can be seen from the implementation strategy, the algorithm is in
essence a series of linear operations -- projections, pixel-rescalings
and convolutions -- that have to be applied step by step to the
pixelised input data. We would like to point out here that for the
derivation of this algorithm we can equally imagine the data to be
gridded on spherical shells in a spherical coordinate system; Fourier
modes would be spherical harmonics on the unity sphere instead of
waves on a plane. In fact, by doing so we will get exactly the same
algorithm that, however, now tells us to linearly combine different
lens-shells or to perform convolutions on the unity sphere. Therefore,
the recipe described in Sect. \ref{sect:kdtree} is a recipe for a
full-sky reconstruction as well, for which the discussed optimisations
and biases apply too.

We also gave an analytic estimate for the noise covariance of density
pixels in the reconstruction, Eq. \Ref{eq:pixelnoise}, for an
idealised survey with homogeneous noise, no gaps and infinite
extent. Several numerical algorithms for the covariance of a realistic
survey are outlined in Appendix \ref{sect:covariance}. A comparison of
the noise to the signal of a SIS, Eq. \Ref{eq:pixelsignalcirc}, shows
that the signal-to-noise of a matter halo quickly decreases towards
increasing redshifts in a reconstruction
(Fig. \ref{fig:sissignaltonoise}).

We noted that the algorithm (if implemented on a flat sky), as other
similar reconstruction methods known in lensing, implicitly assumes
that all cosmic shear signal originates from matter fluctuations
inside the reconstruction area. If a shear of galaxy images inside the
patch actually stems from a source outside the patch, this will
introduce ghost images near the edges of the patch. In principle, the
generation of significant ghost images can be detected by significant
features in the mass density reconstruction based on the B-modes of
the shear field, which come along with the E-mode ghost structures,
and have the same amplitudes. These edge effects can be reduced by
putting an additional frame around the reconstruction patch with
infinite noise.

A fiducial survey with $30$ sources per ${\rm arcmin}^2$, mean
redshift of $\bar{z}=0.85$, $\sigma_\epsilon=0.3$ is unable ($\rm
S/N\lesssim3$) to identify a matter halo with mass
$\sim7\times10^{14}\,h^{-1}\msol$ or less beyond redshift $z\sim0.6$
in a reconstruction with moderate ($\alpha\sim10^{-2}$, transverse
filter) $z$-shift bias. \changed{At a lower redshift of
\mbox{$z\sim0.1$} the detection limit corresponds to a mass of
$\sim1\times10^{14}\,h^{-1}\msol$.}

This redshift limit increases for deeper surveys which we demonstrated
by doing a reconstruction of a mock survey that broadly mimics the
characteristics of a future space-based lensing survey. Still, even
boosting the number density of sources to $100\,\rm arcmin^{-2}$ and
going deeper to $\bar{z}=1.1$ only moderately improves the performance
of the reconstruction. For such a survey, structures in the map of
$\rm S/N\approx 4$ or greater have accurate redshifts within $\pm0.1$,
below this level we have to expect false positives. Generally, we have
to expect to miss out on higher density peaks, even for a powerful
lensing survey as the one assumed here. 

\high{The situation can be expected to become better if a full-sky
reconstruction with heavy smoothing is done to study -- in contrast to
structures on non-linear scales -- density modes on very large scales:
the signal-to-noise increases towards smaller $\ell$, the bias effects
in the reconstruction ease.} \changed{We expect the inclusion of
higher-order lensing distortions into the 3-D reconstruction to
further improve the reconstruction, very roughly increasing the
effective number of ``cosmic shear sources'' by a factor of $N+1$ for
$N$ extra distortion moments. This, however, needs to be explored in
more detail elsewhere.}

\changed{The authors see cosmography as main application of the
reconstruction technique. With 3-D lensing conventional 2-D lensing
convergence maps can be supplemented to some degree by redshift
information, mainly restricted to larger mass overdensities or to
large density modes. This may not give us the full 3-D appearance of
the mass distribution of, say, individual galaxy clusters -- rather
the 2-D projection stretched out in radial direction about the
maximum-likelihood radial distance -- but may enable us to disentangle
physically unconnected structures along the line-of-sight providing a
better understanding of the matter distribution inside a 3-D
volume. This can be vital for a more sophisticated modelling later
on.}

\changed{As an obvious application, the 3-D data cubes may be used to
make catalogues of matter overdensities endowed with redshifts without
any assumptions about density profiles. Estimating the density
profiles of individual structures, essentially projected to 2-D, in
the 3-D lensing map is conceivable, too, but only if the full
reconstruction PSF, including the transverse direction, is accounted
for. Owing to the map biases an estimate for the mass of structures
should be obtained by directly fitting a density model to the shear
data once they have been identified in the non-parametric 3-D
reconstruction.}

\changed{In principle the 3-D maps can be compared to the 3-D
distribution of galaxies as derived from galaxy redshift surveys. In
this case, again, care has to be taken for the biases and limited
radial resolution in the lensing map. For a fair comparison, this may
be done by convolving the map of galaxies with the radial PSF of the
lensing maps. For a statistical comparison of galaxy and (lensing)
matter maps required for studying the galaxy biasing
\citep[e.g.][]{2007A&A...461..861S}, however, we would suggest to
forward and fit a 3-D model of matter density fluctuations and galaxy
biasing to the 3-D shear tomography and its cross-correlation with
galaxy positions, rather than producing a biased Wiener filtered 3-D
matter map that is cross-correlated with the 3-D distribution of
galaxies. In the best case, with all biases in the 3-D mass map being
fully understood, both ways should give same results anyway.}

\changed{The same holds for attempts to estimate matter density power
spectra from the 3-D lensing maps. As shown in this paper, due to low
signal-to-noise the power spectra on the lens planes will be strongly
influenced by the Wiener prior which has to be taken into
consideration for an estimate of the density fluctuation power. Even
if this is done properly, it should only give a result equivalent to
fitting a 3-D model for the shear tomography correlation functions
which is straightforward in comparison \citep{2007ApJS..172..239M}.}

%-------------------------------------------------------------------
\section*{Acknowledgements}
 We would like to thank Alan Heavens, David Wittman, Torsten
 En{\ss}lin, Peter Schneider, Richard Massey and Catherine Heymans for
 useful discussions. We also would like to thank Rachel McInnes for
 helping us with \texttt{MatLab}. PS acknowledges supported by the
 European DUEL Research-Training Network (MRTN-CT-2006-036133) and
 STFC. JH acknowledges support by the Deutsche Forschungsgemeinschaft
 within the Priority Programme 1177 under the project SCHN 342/6 and
 by the Bonn-Cologne Graduate School of Physics and Astronomy.

%-------------------------------------------------------------------
\bibliographystyle{mn2e}
\bibliography{3dmassrec}
% ------------------------------------------------------------------

\appendix

\section{Optimised implementations of the algorithm}
\label{sect:implementation}

\subsection{Real space reconstruction}
\label{sect:kdtree}

As seen in Sect. \ref{sect:densityest}, finding the MV-solution for
some given data vector of shear tomography ellipticities requires, in
the case of the density contrast, the evaluation of
\begin{equation}\label{estimator}
   \vec{\delta}_{\rm MV}= [\mat{1}\alpha+
   \mat{S}_\delta\mat{Q}^{\rm
   t}\mat{P}_{\gamma\kappa}^\dagger\mat{N}^{-1}
   \mat{P}_{\gamma\kappa}\mat{Q}]^{-1}
   \mat{S}_\delta\mat{Q}^{\rm t}
   \mat{P}_{\gamma\kappa}^\dagger\mat{N}^{-1}\vec{\gamma}\;.
\end{equation}
This filter may be used for a reconstruction.  As pointed out by
\citet{1997ApJ...480L..87T}, the estimator Eq. \Ref{eq:shakatoon} is
identical ($\alpha\ne0$) to
\begin{equation}\label{ref:alternative}
  \vec{s}_{\rm
  MV}=\mat{S}\mat{R}^\dagger[\mat{R}\mat{S}\mat{R}^\dagger+\alpha\mat{N}]^{-1}\vec{d}\;,
\end{equation}
which is a form that is also often encountered in the literature. For
that reason, we can equally write as linear filter for the density
contrast field as:
\begin{eqnarray}\label{eq:deltaestimator2}
 &&\!\!\!\!\!\!\!\!\! \vec{\delta}_{\rm MV}=\\\nonumber
 && \!\!\!\!\!\!\!\underbrace{\mat{S}_\delta\mat{Q}^{\rm
  t}\mat{P}_{\gamma\kappa}^\dagger}_{\rm step~3} \underbrace{[
  \mat{N}_{\rm d}^{-1}\mat{P}_{\gamma\kappa}\mat{Q}\mat{S}_\delta\mat{Q}^{\rm
  t}\mat{P}_{\gamma\kappa}^\dagger+\alpha\mat{N}_{\rm
 d}^{-1}\mat{N}_{\rm o}+\alpha\mat{1}]^{-1}}_{=\mat{M}^{-1}, \rm step~2}
    \underbrace{\mat{N}_{\rm d}^{-1}\vec{\gamma}}_{\rm
  step~1}\;.
\end{eqnarray}
Casting the estimator into this form is actually more convenient from
a numerical point of view if we have to consider a non-diagonal noise
covariance, $\mat{N}$, e.g. in the case of intrinsic alignments. Here,
we split the noise covariance, $\mat{N}=\mat{N}_{\rm d}+\mat{N}_{\rm
o}$, into the sum of diagonal, $\mat{N}_{\rm d}$, and off-diagonal,
$\mat{N}_{\rm o}$, elements; the diagonal matrix $\mat{N}_{\rm d}$ is
easily inverted and multiplied with a vector.

For an implementation of this estimator we proceed in three steps. For
the first step, see the underbraces, the binned shear data $\gamma$ is
multiplied by the inverse of the diagonal elements of $\mat{N}$. This
yields the intermediate result ``$\vec{x}$'' that is actually a set of
shear grids in which each grid point is weighed by its noise
variance. For the second step, $\vec{x}$ has to be multiplied by the
matrix $\mat{M}^{-1}$. We employ the method of conjugate gradients to
invert the equation \mbox{$\mat{M}\vec{y}=\vec{x}$} with respect to
$\vec{y}$ \citep {1992nrca.book.....P}. This essentially boils the
matrix inversion down to a series of multiplications with $\mat{M}$
until the solution converges. Again, multiplying by the matrix
$\mat{M}$ is treated as a one-by-one multiplication by the series of
individual matrices contained within the square brackets. As initial
guess for $\vec{y}$ for the iterative algorithm one commonly uses
$\vec{y}=\vec{0}$ which appears to work fine in the context of this
paper. Finally, for the last step $\vec{y}$ is multiplied by
$\mat{S}_\delta\mat{Q}^{\rm t}\mat{P}_{\gamma\kappa}^{\rm t}$.

Multiplying a matrix with a vector means the application of a linear
operation on the input vector. Here, each matrix has a different
effect on the vector that is hard-wired as vector-in/vector-out
sub-routine individually. The multiplication with $\mat{Q}$ (or its
transpose $\mat{Q}^{\rm t}$) is easily implemented as it is just
summing $\sum_iQ_{ij}x^{(i)}_k$ along every line-of-sight
$\vec{\theta}_k$. Multiplications with inverses of diagonal matrices
are also easy as mentioned above. 

The time consuming matrix operations are connected to
$\mat{P}_{\gamma\kappa}$, $\mat{S}_\delta$ and $\mat{N}_{\rm o}$, if
considered, which denote convolutions of the input vector. Only in the
situation where we are using radial Wiener filtering,
$\mat{S}_\delta\vec{x}$ will become an easy operation requiring only
sums along line-of-sights analogous to $\mat{Q}\vec{x}$. The most
effective way of doing convolutions uses Fast-Fourier-Transforms
(FFTs) as an efficient application of the convolution theorem
\citep{1992nrca.book.....P}. For that we need to FFT the different
layers of $\vec{x}$ (either lens planes or source bins) separately and
multiply the Fourier coefficients with the power spectrum of
$\mat{S}_\delta$, the Kaiser-Squires kernel $D(\ell)$, in case of
$\mat{P}_{\gamma\kappa}$, or the power of intrinsic alignments; the
power spectra are functions of lens planes or source bins. The window
function of the grid cells has to be taken into account as well. The
transpose of $\mat{P}_{\gamma\kappa}$ corresponds to
$D^\ast(\ell)$. After that we re-FFT to real space. Note that the
whole chain of operations
\mbox{$\mat{P}_{\gamma\kappa}\mat{Q}\mat{S}_\delta\mat{Q}^{\rm
t}\mat{P}_{\gamma\kappa}^{\rm t}$} can be done completely in Fourier
space with no back and forth FFTing in between.

The FFT method has two drawbacks: i) It requires grids sizes which are
powers of two so that not any arbitrary field-geometry can be
processed. ii) Also, as well known, we will have edge effects as FFTs
are assuming periodic boundary conditions (aliasing). Regarding i) we
have to find the smallest bounding box to enclose the whole field of
view at the desired resolution. Resolutionwise a number of grid cells
of $128$ or $64$ for each coordinate axis is usually enough. With
respect to ii) we can use the zero-padding technique or down-weighting
towards the edges (see next section) to mitigate the aliasing.

There is an alternative technique known in the literature that speeds
up convolutions but works solely in real space
\citep{2003NewA....8..581P}, does not suffer from FFT-aliasing and the
geometry constraint and, in fact, does not require griding. This
approach is a real alternative to FFT but considerably slower as we
found. As both techniques yield essentially the same reconstructions,
apart from the edges, we propose to stick to the FFT technique in this
context.

Processing $25$ lens planes and $40$ source redshift bins, all with a
grid size of \mbox{$128\times128\,\rm pixel^2$} (the data vector,
complex, and matter density vector, complex (E-, B-mode combined),
have sizes of $1.3\times10^6$ and $8.2\times10^5$ elements,
respectively) requires for the algorithm $\sim60\,\rm MByte$ of
computer memory (double precession). The computation time for the
second step on a AMD Athlon 64 processor with 2.4GHz is approximately
five seconds for every iteration of step 2, which is the main
bottleneck of the algorithm. For noisy data and (full,
i.e. $\alpha=1$) Wiener filtering usually about five iteration steps
are needed, whereas a reconstruction with no prior or heavily
tuned-down Wiener filter ($\alpha\sim0.01$) requires up to $\sim150$
iteration steps, depending on the desired accuracy, because less
regularisation is applied to the data so that more details are
fitted. The benchmark parameters used here are usually, with
contemporary real data, unnecessary. For that $64^2$-grids, $N_{\rm
lp}\sim 10$ lens planes and $N_{\rm z}\sim 20$ source bins are
absolutely sufficient bringing down the computation time of one full
reconstruction to roughly ten seconds.

After the algorithm has converged, we run a final (transverse)
smoothing of the lens-plane grids with a Gaussian kernel. The
smoothing is usually required as a down-tuned Wiener filter is applied
that produces noisy images. The blurring can be done efficiently by
again employing the FFT convolution or a kdtree implementation.

\subsection{Fourier space reconstruction}
\label{sect:fourierrecon}

If an idealised survey is a good assumption for particular data, the
Fourier space filter Eq. \Ref{estimatorfourier} is, due to the small
matrices involved, a very fast and efficient way to perform a
3-D reconstruction completely in Fourier space. In particular, the
matrix inversions can be done explicitly. To apply this filter, we FFT
the gridded source $z$-bins, arrange the shear Fourier coefficients
for each $\vec{\ell}$ inside one vector, $\tilde{\vec{\gamma}}$, and
perform Eq. \Ref{estimatorfourier} for every angular mode. The
matrices involved have only sizes of the order of $N_{\rm z}$ or/and
$N_{\rm lp}$, which are easily and quickly inverted. The Fourier
reconstruction is eventually back-FFTed to acquire the real space mass
density field. Finally, the lens-planes need to be smoothed, since we
are usually using a tuned down Wiener filter ($\alpha=0.01$, for
instance). This smoothing can be done in real space or, more
efficiently, directly in Fourier space immediately after the Fourier
space reconstruction.

Usually for $\ell\ne0$ a regularisation ($\lambda\ne0$), especially
for the no-prior MV-estimator, is not necessary in a Fourier space
reconstruction because the reason for the singularity of the filter
is the mass-sheet degeneracy that affects solely the $\ell=0$
modes. Here, we lift this degeneracy by setting all $\ell=0$ Fourier
coefficients to zero which amounts to saying that the mean density
contrast on each lens plane is exactly zero. For a large enough
field-of-view this should be a reasonable assumption.

In order to reduce the FFT edge effects, the real space grids should,
initially, be downscaled by $\sin{\left(\pi\Delta/2\Delta_0\right)}$
within a strip of width $\Delta_0$, where $\Delta$ is the shortest
distance of a pixel to the grid edge.

The noise covariance is estimated by means of the well-known
shot-noise formula $[\sigma^{(i)}_\epsilon]^2/\bar{n}_i$
(Sect. \ref{sect:densitycontrast}), ideally assuming that the grid
cells are homogeneously filled with sources of roughly equal intrinsic
ellipticity distribution.

A complete reconstruction on a Fourier grid for the same parameters
(but $128\times128$ grids) as in the foregoing section takes less than
a second.

\subsection{Covariance of statistical uncertainties}
\label{sect:covariance}

An estimation of the statistical errors, or more general the
covariance, of pixel densities in the reconstruction is needed to
convert the density reconstruction into a signal-to-noise 3-D map. As a
lowest-order estimate we suggest to use the analytical solution,
Eq. \Ref{eq:pixelnoise}, that is based on homogeneous noise of the
sources. It offers the r.m.s.-error for each lens-plane and the
correlation of errors along each line-of-sight.

As a final smoothing of each lens-plane will usually be necessary, the
estimate has to be rectified slightly. Lets us assume that the errors,
$\sigma_\delta$, of the unsmoothed map are uncorrelated, which will be
roughly correct for a strongly down-tuned (small $\alpha$) filter. By
smoothing a lens-plane with a normalised kernel, $K(x)$, one will
combine an effective number of
\begin{equation}
  N_{\rm eff}\equiv
  \frac{\left[\int_{R^2}\d^2\vec{\theta}K(|\vec{\theta}|)\right]^2}
       {\int_{R^2}\d^2\vec{\theta}[K(|\vec{\theta}|)]^2}
\end{equation}
pixels with uncorrelated errors $\sigma_\delta$. The combined error of
the weighted average (every smoothed pixel) will therefore be
$\sigma_\delta/\sqrt{N_{\rm eff}}$. For a Gaussian kernel with width
$\sigma$ one finds $N_{\rm eff}=\pi^2\sigma^2$.  The truncation of the
kernel due to the edges of the map can be accounted for using the
formula for $N_{\rm eff}$ and setting $K(x)=0$ for areas outside the
map. As the pixels one is smoothed over are slightly correlated,
$\sigma_\delta/\sqrt{N_{\rm eff}}$ will give only a lower limit for
the true error (upper limit for the signal-to-noise), whereas
$\sigma_\delta$ would be the noise upper limit being reached if all
pixels were $100\%$ correlated.

In reality, one encounters surveys with varying source number
densities and intrinsic source shot noise. One may desire to get a
better estimate for the signal-to-noise of a pixel than provided by
the foregoing formula, which is given for all pixels by
\begin{equation}
  \mat{N}_{\rm MV}=\mat{W}\mat{X}^{-1}\,\mat{W}^\dagger\;,
\end{equation}
where \mbox{$\mat{X}^{-1}=\mat{Q}^{\rm
    t}\mat{P}^\dagger_{\gamma\kappa}\mat{N}^{-1}\mat{P}_{\gamma\kappa}\mat{Q}$},
$\mat{W}=\mat{F}^\dagger\mat{S}_\delta[\mat{X}^{-1}\mat{S}_\delta+\alpha\mat{1}]^{-1}$
and $\mat{F}$ is the smoothing matrix (unity matrix in absence of
final smoothing).  However, the full covariance imposes a huge, if
not almost impossible, computational task because of the enormous
size of the involved matrices.

What can easily be done, though, is to compute single matrix elements
of the covariance:
\begin{eqnarray}\label{cov1}
  [\mat{N}_{\rm MV}]_{ij}&=&
  \vec{v}_i^\dagger\mat{N}^{-1}\vec{v}_j\;,\\
  v_i&\equiv&\mat{P}_{\gamma\kappa}\mat{Q}[\mat{S}_\delta
    \mat{X}^{-1}+\alpha\mat{1}]^{-1}\mat{S}_\delta\mat{F}_j\;,
\end{eqnarray}
where $\mat{F}_i$ is defined as the $i$th column of the matrix
$\mat{F}$, which is the smoothing weight of any grid pixel relative
to a fixed grid pixel $i$. This vector is filled with many zero
because we are only smoothing within the same lens-plane. As
indicated, the recipe is to compute the vectors $\vec{v}_{i/j}$ in a
first step and then \mbox{$\vec{v}_i^\dagger\mat{N}^{-1}\vec{v}_j$}
in a second step. In order to attach error bars to single
$\delta_{\rm MV}$ one requires only one vector
$\vec{v}_i=\vec{v}_j$.

The definition of $\vec{v}_{i/j}$ is almost identical to $\delta_{\rm
MV}$ in Eq. \Ref{estimator} so that the algorithm discussed in
Sect. \ref{sect:kdtree} applies without much change. On the other
hand, this means the effort to obtain a matrix element of
$\mat{N}_{\rm MV}$ is roughly twice the effort (Only off-diagonals;
the same effort for diagonals) to remake a whole 3-D reconstruction, so
we may only wish to work out the exact signal-to-noise or correlation
of errors for a selected small number of pixels such as the peaks in
the mass map that are suspected to belong to galaxy clusters.

A third and most effective alternative for obtaining a signal-to-noise
map is to randomise the ellipticities of the sources and to make a
full reconstruction.  With the algorithm outlined in
Sect. \ref{sect:kdtree} this can be done hundreds of times in a
reasonable time, maybe running parallelly on different computers. The
variance in the density contrast between the different noise
realisations would serve as estimator for the noise level in the
reconstruction.

\section{Shear-intrinsic alignment correlations}
\label{sect:shearintr}

The estimator Eq. \ref{ref:alternative} is the minimum variance
solution, or the most likely solution for $\vec{s}$ in the Bayesian
sense for pure Gaussian statistics, for a given data vector $\vec{d}$
provided that there is no correlation between noise and signal,
i.e. $\ave{\vec{s}\vec{n}^\dagger}=\mat{0}$. In the context of
gravitational lensing, however, these correlations can occur if, for
example, the shear signal itself is correlated to the intrinsic
alignment of sources \citep{2004PhRvD..70f3526H}. We ignore this
effect for 3-D mass reconstructions in this paper but would like to
sketch how it can, in principle, be included by extra terms in the
minimum-variance estimator for $\vec{s}$.

For that we assume that the signal/noise-correlations (here:
correlations between the lens plane matter density and the intrinsic
ellipticities of sources) between the various source $z$-bins are
known either from measurement or from a physical model,
$\mat{U}\equiv\ave{\vec{s}\vec{n}^\dagger}$.  The minimum-variance
Ansatz seeks to find a linear transformation, $\mat{H}$, that
minimises the residual matrix
\begin{equation}\label{ref:minvar}
  \Ave{(\mat{H}\vec{d}-\vec{s})(\mat{H}\vec{d}-\vec{s})^\dagger}
\end{equation}
for all possible noise and signal configurations. For the simpler case
that $\mat{U}$ vanishes, this residual matrix is, leaving out terms
independent of $\mat{H}$, on average 
\begin{equation}\label{ref:variance}
  \mat{H}(\mat{R}\mat{S}\mat{R}^\dagger+\mat{N})\mat{H}^\dagger-\mat{H}\mat{R}\mat{S}-\mat{S}\mat{R}^\dagger\mat{H}^\dagger\;,
\end{equation}
which is minimised for
\begin{equation}\label{ref:H1}
  \mat{H}=\mat{S}\mat{R}^\dagger[\mat{R}\mat{S}\mat{R}^\dagger+\mat{N}]^{-1}\;.
\end{equation}
If we now drop the condition of a vanishing $\mat{U}$, the variance
\Ref{ref:minvar} becomes more generally
\begin{equation}
  \mat{H}(\mat{R}\mat{S}\mat{R}^\dagger+\mat{N}+\mat{R}\mat{U}+\mat{U}^\dagger\mat{R}^\dagger)\mat{H}^\dagger-\mat{H}(\mat{R}\mat{S}+\mat{U}^\dagger)-(\mat{S}\mat{R}^\dagger+\mat{U})\mat{H}^\dagger\;.
\end{equation}
If we substitute in the last expression
\begin{eqnarray}
  \mat{R}\mat{S}&\mapsto&\mat{R}\mat{S}-\mat{U}^\dagger\;,\\
  \mat{N}&\mapsto&\mat{N}-\mat{R}\mat{U}\;,
\end{eqnarray}
we recover exactly the variance \Ref{ref:variance} with the already
known minimising solution Eq. \Ref{ref:H1}. Undoing the former
substitution in \Ref{ref:H1} (the substitution is independent of
$\mat{H}$) gives the more general minimum-variance that now also
accounts for signal-noise correlations:
\begin{equation}
  \mat{H}=\left(\mat{S}\mat{R}^\dagger+\sqrt{\alpha}\mat{U}\right)
  \left[\mat{R}\mat{S}\mat{R}^\dagger+\alpha\mat{N}+\sqrt{\alpha}\mat{R}\mat{U}+
  \sqrt{\alpha}\mat{U}^\dagger\mat{R}^\dagger\right]^{-1}\;.
\end{equation}
Here, we already included the rescaling of
\mbox{$\mat{S}\mapsto\alpha^{-1}\mat{S}$} needed for the
$\alpha$-tuning (Saskatoon filter) supplemented by
\mbox{$\mat{U}\mapsto\sqrt{\alpha}^{-1}\mat{U}$} to be consistent with
the $\alpha$-tuning (the signal-noise correlations are unchanged by
changing $\alpha$). We find that including the signal-noise
correlation adds three new terms to Eq. \Ref{ref:H1}. In practice,
this estimator can be tackled by the same numerical techniques as
outlined in Appendix \ref{sect:implementation}.

\bsp

\label{lastpage}

\end{document}